\documentclass[structabstract]{aa}
\usepackage{graphicx}
\usepackage{txfonts}
\usepackage{color}
\usepackage{lscape}
\usepackage{natbib}

\newcommand{\kms}{km~s$^{-1}$}

\begin{document}

\newcommand{\gsim}{\raisebox{-.4ex}{$\stackrel{>}{\scriptstyle \sim}$}}
\newcommand{\lsim}{\raisebox{-.4ex}{$\stackrel{<}{\scriptstyle \sim}$}}
\newcommand{\psim}{\raisebox{-.4ex}{$\stackrel{\propto}{\scriptstyle \sim}$}}
\newcommand{\doce}{\mbox{$^{12}$CO}}
\newcommand{\trece}{\mbox{$^{13}$CO}}
\newcommand{\jcc}{\mbox{$J$=5$-$4}}
\newcommand{\jdu}{\mbox{$J$=2$-$1}}
\newcommand{\juc}{\mbox{$J$=1$-$0}}
\newcommand{\mloss}{\mbox{$\dot{M}$}}
\newcommand{\my}{\mbox{$M_{\odot}$~yr$^{-1}$}}
\newcommand{\ls}{\mbox{$L_{\odot}$}}
\newcommand{\ms}{\mbox{$M_{\odot}$}}
\newcommand{\mm}{\mbox{$\mu$m}}
\def\arcdeg{\hbox{$^\circ$}}
\newcommand{\secp}{\mbox{\rlap{.}$''$}}

\title{SHAPEMOL: a 3-D code for calculating CO line emission in planetary and
protoplanetary nebulae}

\subtitle{Detailed model-fitting of the complex nebula NGC 6302}

\author{M. Santander-Garc\'\i a\inst{1,2} %,3,4}
\and V. Bujarrabal\inst{1}
\and N. Koning\inst{3}
\and W. Steffen\inst{4}
%\and someone else?
}

\institute{
Observatorio Astron\'omico Nacional, Ap.\ de Correos 112, E-28803, Alcal\'a de Henares, Madrid, Spain \\ email: {\tt m.santander@oan.es}
\and
Instituto de Ciencia de Materiales de Madrid (CSIC), E-28049, Madrid, Spain
\and
Department of Physics \& Astronomy, University of Calgary, Calgary, Alberta, Canada
\and
Instituto de Astronom\'\i a Universidad Nacional Aut\'onoma de M\'exico, C.P. 22860, Ensenada, Mexico
}

% \date{Received January 6, 2010; accepted March 16, 2026}

% \abstract{}{}{}{}{}
% 5 {} token are mandatory

\abstract
{Modern instrumentation in radioastronomy constitutes a valuable tool for studying the Universe: ALMA has reached unprecedented sensitivities and spatial resolution, while Herschel/HIFI has opened a new window (most of the sub-mm and far-infrared ranges are only accessible from space) for probing molecular warm gas ($\sim$50-1000~K). On the other hand, the software {{\tt SHAPE}}  has emerged in the past few years as a standard tool for determining the morphology and velocity field of different kinds of gaseous emission nebulae via spatio-kinematical modelling. Standard {{\tt SHAPE}} implements radiative transfer solving, but it is only available for atomic species and not for molecules.}
{Being aware of the growing importance of the development of tools for simplifying the analyses of molecular data from new-era observatories, we introduce the computer code {{\tt shapemol}}, a complement to {{\tt SHAPE}}, with which we intend to fill the so-far under-developed molecular niche.}
{{{\tt shapemol}} enables user-friendly, spatio-kinematic modelling with accurate non-LTE calculations of excitation and radiative transfer in CO lines. Currently, it allows radiative transfer solving in the $^{12}$CO and $^{13}$CO $J$=1$-$0 to $J$=17$-$16 lines, but its implementation permits easily extending the code to different transitions and other molecular species, either by the code developers or by the user.  Used along {{\tt SHAPE}}, {{\tt shapemol}} allows easily generating synthetic maps to test against interferometric observations, as well as synthetic line profiles to match single-dish observations.}
{We give a full description of how {{\tt shapemol}} works, and we discuss its limitations and the sources of uncertainty to be expected in the final synthetic profiles or maps. As an example of the power and versatility of {\tt shapemol}, we build a model of the molecular envelope of the planetary nebula NGC 6302 and compare it with \doce\ and \trece\ \jdu\ interferometric maps from SMA and high-$J$ transitions from Herschel/HIFI. We find the molecular envelope to have a complex, broken ring-like structure with an inner, hotter region and several  `fingers' and high-velocity blobs, emerging outwards from the plane of the ring. We derive a mass of  0.11 M$_\odot$ for the molecular envelope.}
{}


\titlerunning{SHAPEMOL: a 3-D code for calculating CO line emission in PNe and PPNe}

\maketitle
%
%________________________________________________________________

\section{Introduction }

Radioastronomical instrumentation has undergone significant improvement in recent years, leading to notable advances in our understanding of the Universe at molecular wavelengths. On the one hand, the spatial resolution achievable with the Atacama Large Millimeter/submillimeter Array (ALMA) rivals that of the Hubble Space Telescope in the optical regime, with unprecedented sensitivities. On the other hand, the Heterodyne Instrument for the Far Infrared aboard Herschel (HIFI) is an invaluable tool as it has effectively opened a new window from which to probe warm molecular gas ($\sim$50--1000 K). While incapable of generating images of the source because of its low spatial resolution, and thus lacking information on the spatial distribution of its molecular gas, HIFI produced 1-D high-resolution spectra in the range of 480-1250~GHz and 1410-1910~GHz, which includes transitions as high as $^{12}$CO and $^{13}$CO $J$=16$-$15, unobservable from the ground ---as is most of the sub-mm and far-infrared range. In operation until April 2013, Herschel+HIFI has contributed to the study of the excitation conditions of the warm molecular gas of many astrophysical nebulae (e.g. \citealp{bujarrabal11}).

%With its 54 12-m antennas with baselines up to 16 km, together with an array of 12 7-m antennas, ALMA is able to reach spatial resolutions as high as 0\farcss 006 at 675 GHz and 0\farcss 038  at 110 GHz, and with unprecedented sensitivities. 

%See, as an example of the ALMA capabilities during the early science phase, the recently published cycle 0 data by \cite{boley12}.

%As of today, in cycle 1 and with 41 antennas in place (baselines up to $\sim$1~km), ALMA is capable of reaching a spatial resolution of 0\farcss 08 in band 9 (e.g. $^{12}$CO and $^{13}$CO $J$=6$-$5 transitions)

The study of nebulae around evolved stars, where molecular gas is very common, can greatly benefit from the power of modern radioastronomical instrumentation, such as ALMA's spatial resolution and sensitivity and the ability of Herschel+HIFI to probe warm molecular gas. Being aware of the growing importance of the development of tools for simplifying the analyses of molecular data from new-era observatories, we introduce the computer code {{\tt shapemol}}. This complementary software enhances the ability of {\tt SHAPE}, the often-used tool for performing spatio-kinematical modelling in the field of protoplanetary (PPNe) and planetary nebulae (PNe), by allowing the study of the excitation conditions of nebulae from the analysis of CO lines, which are known to be the best tracers of molecule-rich gas in these objects. Together, {\tt SHAPE}+{\tt shapemol} can generate high-resolution synthetic spectral profiles and maps to be compared with observational data. Although {\tt SHAPE}+{\tt shapemol} are primarily intended for the study of PPNe and PNe, any other astrophysical nebulae that fulfill some general conditions (see~\ref{treatment}) can be studied with this software. {\tt shapemol} has been succesfully tested in the study of the excitation conditions of the molecular envelope of the planetary nebula NGC~7027 using data from Herschel/HIFI and IRAM 30m (\citealp{santander12}), and is applied here to the molecular envelope of the young PNe NGC 6302, showing the power of {\tt shapemol} and yielding an accurate description of the complex structure and physical conditions of its molecular envelope.

The present work is structured as follows: We provide a brief description of {\tt SHAPE} and the motivation for {\tt shapemol} in~\ref{shape}. Section~\ref{shapemol} comprehensively describes this complementary software, including both its treatment of the radiative transfer, based on the well-known large velocity gradient (LVG) approximation, and the implementation of the software itself. Section~\ref{uncertainty} outlines the uncertainties to be expected in the synthetic results, and~\ref{limitations} covers the limitations of {\tt shapemol}. Finally, the science case is presented in section~\ref{science} with the application of {\tt SHAPE}+{\tt shapemol} to the molecular envelope of NGC 6302.

\section{SHAPE}\label{shape}

{\tt SHAPE}\footnote{http://bufadora.astrosen.unam.mx/shape/} is an interactive 3-D software tool for modelling
complex gaseous nebulae (mainly planetary nebulae, but
also supernova remnants, light echoes, emission nebulae from
massive stars, high-energy phenomena, etc). The distribution of density, velocity, and other physical
properties is generated interactively using 3-D mesh structures
and other graphical and mathematical tools. From these data the program
generates synthetic images, position-velocity diagrams, 1-D spectral
profiles, and channel maps for direct comparison with observations.
Its versatility has made it a standard tool for the 3-D reconstruction
of planetary nebula (e.g. \citealp{steffen11}, \citealp{jones10b})
and the analysis of hydrodynamical
simulations (e.g. \citealp{steffen09a}; \citealp{velazquez11}).
{\tt SHAPE} implements radiative transfer solving for atomic
species using coefficients from the CHIANTI (\citealp{landi12}), Kurucz (\citealp{smith96}), and NIST (\citealp{reader12}) databases. However, molecular physics in thermalised and
non-thermalised cases was not implemented in {\tt SHAPE} until now.
We designed {\tt shapemol} to fill this gap.

While an earlier version of {\tt shapemol} worked as a complement to {\tt SHAPE} v4.5, it has been fully integrated into {\tt SHAPE} v5. In its present state, {\tt shapemol} enables radiative transfer in \doce\ and \trece\ lines. This is done by interpolating the absorption
and emission coefficients from a set of pre-generated tables computed
under the assumption of the LVG
approximation. 

This choice of approximate solutions over exact ones reflects the philosophy behind shapemol: while a typical 3-D code with 10$^4$ grid cells would take several minutes to compute the exact solution in current regular computers, {\tt SHAPE}+{\tt shapemol} will handle models consisting of 10$^6$ grid-cells in a matter of 30-40 seconds. This, together with its integration within the user-friendly {\tt SHAPE} software, already a standard in the field, makes {\tt shapemol} a quick and versatile tool for analysing molecular data, as it allows the generation of almost any 3-D structure and velocity fields with a few mouse clicks.

In the next section, we describe the specific treatment of the radiative transfer behind this complementary software as well as its implementation and interaction with {\tt SHAPE}.

\section{SHAPEMOL}\label{shapemol}

{\tt shapemol} produces synthetic spectral profiles and maps of \doce\ and \trece\ lines to be compared with sub-mm and mm-range observations. In particular, {\tt shapemol} computes the absorption and emission coefficients $k_\nu$ and $j_\nu$ for a large number of cells defined within the nebula, as a function of a set of parameters given by the {\tt SHAPE} model (i.e. positions, velocities, densities, and temperature of each point in the nebula) together with additional input (i.e. abundances, desired transition, and species) from the user. {\tt SHAPE} then uses the computed coefficients for solving the radiative transfer equation, generating synthetic spectral profiles, and/or maps of the nebula at the given transition.

{\tt shapemol} is included in {\tt SHAPE} v5.0, although the tables on which it is based (see Sect.~\ref{implementation}) have to be downloaded separately\footnote{http://www.astrosen.unam.mx/shape/v5/Downloads/SHAPE\_INSTALLERS/CO\_tables.zip}. The tables are freely available, the only requisite being that in publications that include models prepared with {\tt shapemol}, credit should be given citing this work (as well as citing the paper describing {\tt SHAPE} by \citealp{steffen11}). 

\subsection{Treatment of the radiative transfer and level population} \label{treatment}

{\tt shapemol} solves the standard radiative transfer equation to
calculate the line intensity along a large number of lines of sight and
many frequencies around the resonant line:
\begin{equation}
{\rm d}I_\nu / {\rm d}l ~=~ k_\nu I_\nu + j_\nu ~~,
\end{equation}
where $k_\nu$ and $j_\nu$ are the absorption and emission coefficients of a given point in the nebula,
which depend on the populations of the levels joined by the transition. We assume that the
frequency dependence of the normalized line profile $\phi_\nu$ is the
same for both coefficients, $k_\nu$ = $k \phi_\nu$, $j_\nu$ = $j
\phi_\nu$. This {\em total redistribution} assumption is justified in
our case and particularly for CO lines because of the high probability
of the elastic collisions and the relatively low radiative
probabilities. In the case of IR and mm-wave transitions from diffuse
media (interstellar or circumstellar clouds), $\phi_\nu$ is always
given by the local velocity dispersion, that is, the thermal or
microturbulent velocities. The (non-LTE) level populations used to
solve the transfer equation are calculated using the well-known large velocity gradient (LVG)
approximation.

The level populations depend on the collisional transition rates and
the radiative excitation and de-excitation rates, which in turn depend
on the amount of radiation reaching the nebula point we are considering
at the frequency of the line (averaged over angle and frequency within
the local line profile). The calculation of this averaged radiation
intensity requires a previous knowledge of the absorption and emission
coefficients in the whole cloud to solve the radiative
transfer equation in all directions and frequencies. Indeed, the
populations of very many levels must be calculated
simultaneously, since the population of each one depends on those of
the others, and in all the points of the cloud. This renders the solution of the
system extremely complex in the general case.

The problem is greatly simplified when there is a large velocity
gradient in the cloud, introducing significant Doppler shifts
between points that are sufficiently far away. When this shift is
larger than the local velocity dispersions, the points cannot
radiatively interact at large scales, so the radiative transfer is
basically a local phenomenon. The interaction of molecules and radiation at a given
point can be described by the well-known {\em
  escape probability}. The excitation equations can then be solved and
$k_\nu$ and $j_\nu$ calculated at each point, independently of the rest
of the cloud, in the frame of the LVG approximation. In any case, the
level populations depend in a complex way on the (local) physical
conditions, and the solution requires an iterative process. See
\cite{castor70} for a comprehensive version of the general LVG
formalism. The LVG approximation includes the main ingredients of the
problem (collisional and radiative rates, trapping when opacities are
high, population transfer between different levels, etc.) and gives
fast, sensible solutions. These excitation calculations are quite
accurate, even when the LVG conditions are barely satisfied,
particularly in the case of AGB and post-AGB shells; see detailed
comparison with non-local {\em exact} calculations in Bujarrabal and
Alcolea (2013), and further use of the LVG formalism by \cite{teyssier06} 
and \cite{ramstedt08}.  The approximation itself is
not necessary to derive the resulting line profiles, which can be
calculated solving the standard, full transfer equation using the level
populations derived from the LVG approximation and a local velocity
dispersion, as indeed we have done using {\tt SHAPE}. The use of the LVG approximation to calculate the absorption and emission coefficients is basic to keep the simple and fast use of {\tt shapemol} in very general 3-D cases, since in this approximation the coefficients only depend on a few local parameters, and not on the general structure of the nebula.

According to the LVG approximation, $k_\nu$ and $j_\nu$ depend in a
heavily non-linear way on the density $n$ and the temperature $T$ of
each point, and almost linearly on the product $\frac{r}{V} X$,
where $r$ is the distance of a given point (or grid cell) to the
central star, $V$ its velocity, and $X$ the abundance of the
species. In addition to these three parameters, the results of the LVG depend
on the logarithmic velocity gradient, $\epsilon$ = d$V$/d$r$\,$r$/$V$,
but only slightly. In its current state, {\tt shapemol} lets the user
select values of $\epsilon$=0.2, 1 and 3. The calculation of the table values are simplified in the most common case, when
$\epsilon$ = 1, that is, a linear dependence of $V$ on $r$, since the
escape probability of a photon in that case is given by a simple
analytical function of the opacity. Such velocity fields are found in
many young planetary nebulae, which are basically expanding at high
velocity following a ``ballistic'' velocity law (i.e. with a linear dependence on the distance from the central star). We stress that the LVG
results (in the standard case) only depend on four parameters: $n$,
$T$, $\frac{r}{V} X$, and $\epsilon$.

In our calculations, we used the most recent collisional rates from the
LAMDA database\footnote{http://www.strw.leidenuniv.nl/$\sim$moldata},
\cite{schoier05}, and rates calculated by \cite{yang10} for
collisions with ortho- and para-H$_2$, for which we assumed an
abundance ratio of 3.  Collisions with other gas components, notably He, are not
considered, so the derived density represents the total density and not
only the density of H$_2$.

%(Results slightly depend on the assumed 
%ratio and we cannot be sure under which excitation conditions H$_2$ was
%formed.)  

Absorption and emission coefficients were obtained for the 17 lowest
rotational transitions of the ground-vibrational state, practically the
only ones observed. In calculating them, a total of 40 rotational
levels were considered, although in low-excitation cases only the
lowest ones are significantly populated. We verified the
convergence of the calculated coefficients on the total number of
levels, which was expected because we only considered temperatures well
below the energies of the highest considered levels.  Except for tests
of the calculation accuracy, we only considered the rotational
transitions in the ground-vibrational state, since vibrational
excitation has been found to have negligible effects in practical
cases, see Appendix A. In fact, calculations including vibrational
excitation are easy to perform (we just have to include more levels and
transitions in the code), but this limitation is important for our
calculations to be reasonably general. Otherwise, IR continuum sources in the
studied nebulae, which cause most vibrational transitions, should
be taken into account, resulting in line excitations that would depend
in a complex way on the structure and intensity of IR sources, on the
distance between them and the considered points, and on the absorption
of IR photons in their trajectory between both. The general-purpose design of {\tt shapemol}, as well as its user-friendliness and fast calculation, would be strongly limited in that case.

As we see in Appendix A, the negligible contribution of the IR
excitation in fact holds in regions with significant emission. When the
lines are very weak, for instance because of a very low density, the
contribution of IR excitation can be important compared with the
emission if IR is not included, but remaining very low in any case. In
practical cases, that is, whenever the lines are detectable, the
contribution of IR excitation to the total emission is very low.

These continuum sources also emit at rotational-line frequencies. In the current version of {\tt shapemol}, the absorption by rotational transitions of photons from the cosmic background and the rotational transitions themselves are taken into account, but not those from the local continuum sources. In our calculations in Appendix A, we take into account the effects of this absorption on the molecular level population and line formation. We have checked that these effects are even lower than those due to vibrational excitation by IR radiation.

We recall that vibrational excitation can be important in other
sources, for instance for high-$J$ emission from circumstellar envelopes around AGB stars under certain conditions,
because in these objects the densities (and therefore the collisional
excitation) are lower and the continuum IR emission is higher.  We finally note that for molecules
other than CO, the vibrational excitation can be more intense, because
carbon monoxide is easily excited by collisions (and its levels
thermalised).

\begin{figure*}[!]
\center
\resizebox{15cm}{!}{\includegraphics{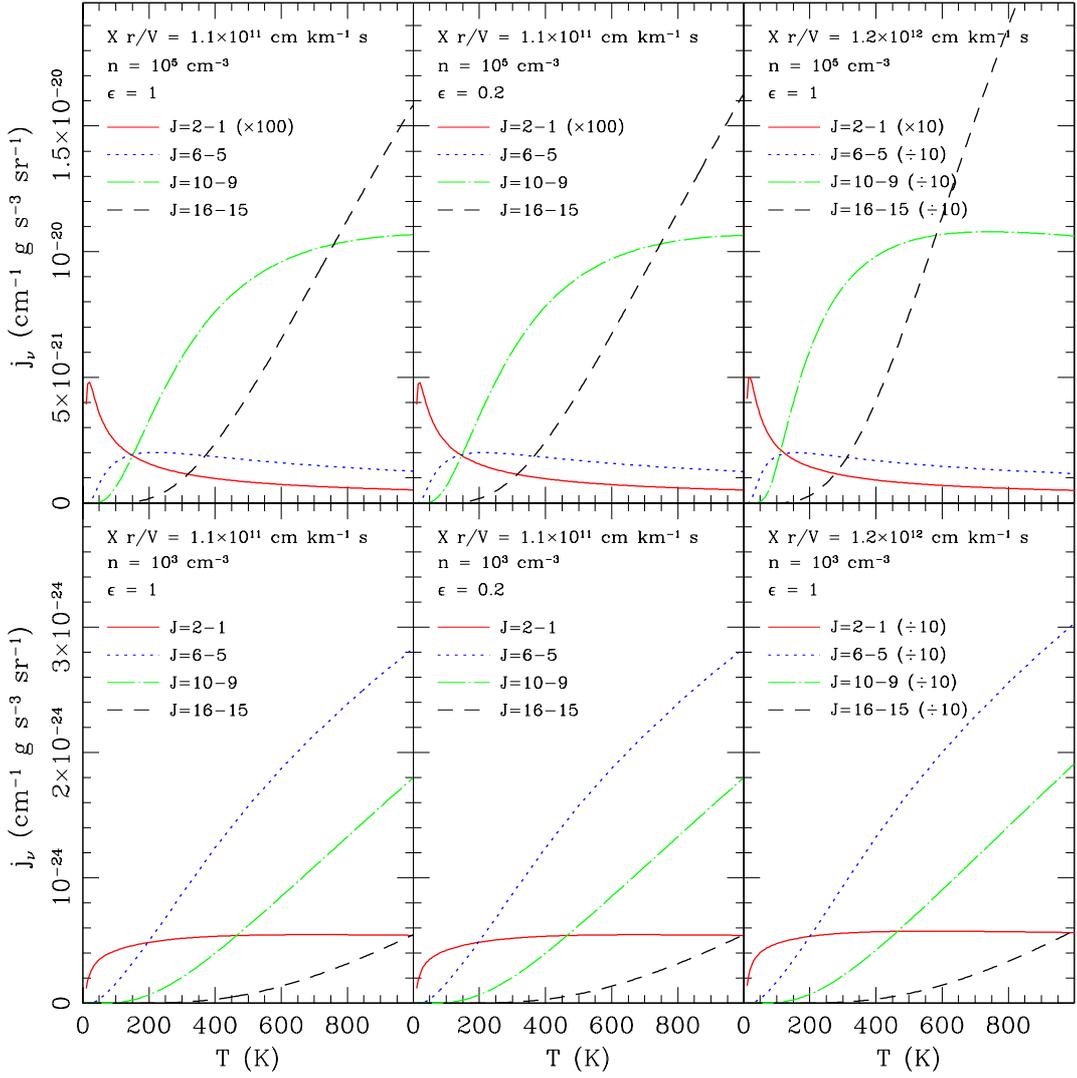}}
\caption{Examples of the $j_\nu$ coefficient corresponding to the $^{12}$CO transitions with $J$=2--1, 6--5, 10--9 and 16--15 as a function of temperature in six different physical conditions, with varying $\epsilon$, $n$ and $\frac{r}{V} X$. Some of the transitions have been scaled up or down by the specified factors for displaying purposes. The non-linear dependance on the density $n$ can easily be seen here, as well as the almost linear dependance on the product $\frac{r}{V} X$. The value of $\epsilon$, on the other hand, affects the $j_\nu$ coefficient only slightly.}
\label{FJ}
\end{figure*}

\begin{figure*}[!]
\center
\resizebox{15cm}{!}{\includegraphics{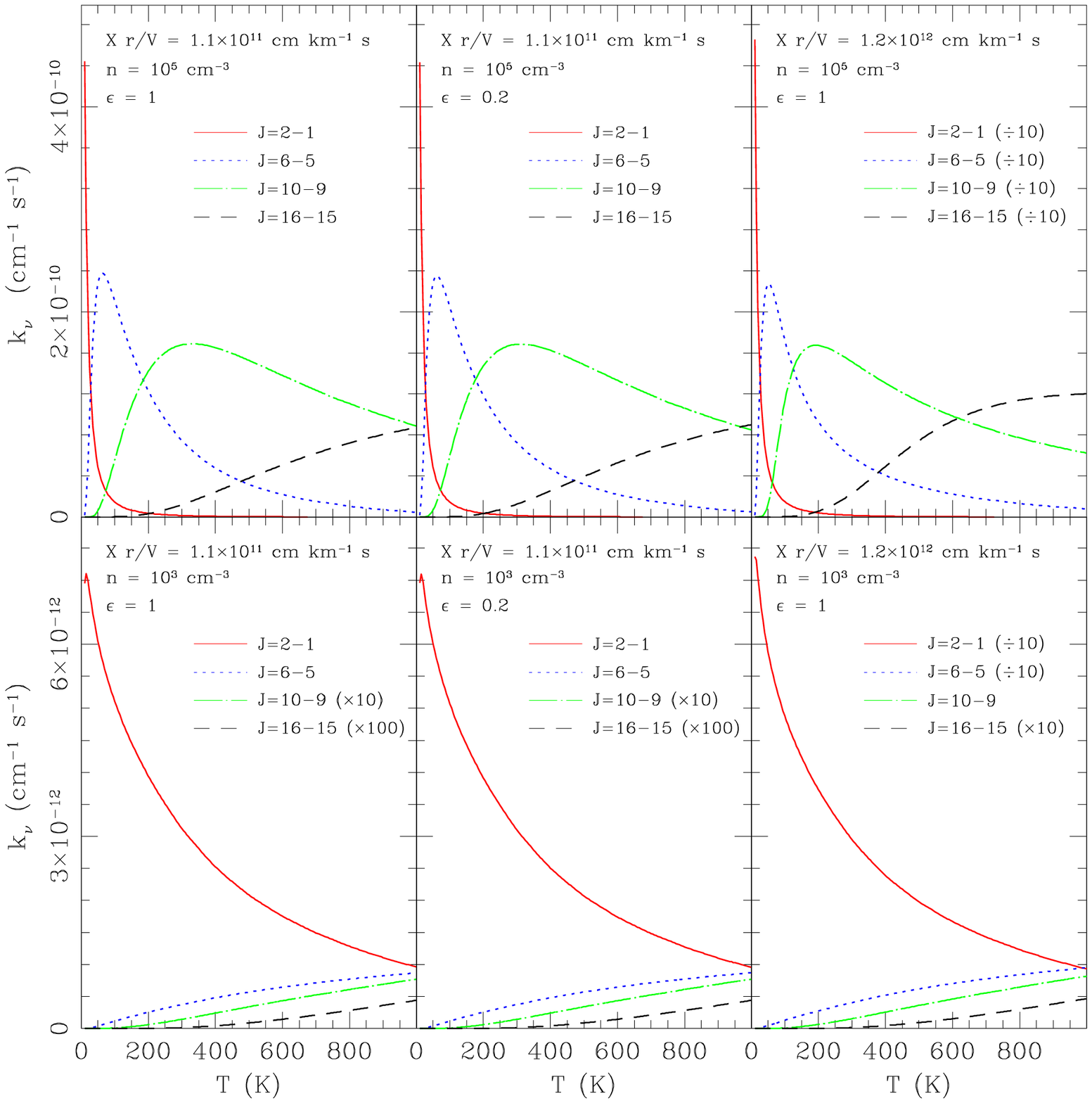}}
\caption{Examples of the $k_\nu$ coefficient corresponding to the $^{12}$CO transitions with $J$=2--1, 6--5, 10--9 and 16--15 as a function of temperature in six different physical conditions, with varying $\epsilon$, $n$ and $\frac{r}{V} X$. Some of the transitions have been scaled up or down by the specified factors for displaying purposes. The non-linear dependance on the density $n$ can easily be seen here, as well as the almost linear dependance on the product $\frac{r}{V} X$. The value of $\epsilon$, on the other hand, affects the $k_\nu$ coefficient only slightly.}
\label{FK}
\end{figure*}

\subsection{Implementation of {\tt shapemol}}\label{implementation}

In principle, computing the $k_\nu$ and $j_\nu$ coefficients for each point of the model would require solving the population level equations once for the physical conditions of each single point via the convergence algorithm typical of LVG codes. In practice, however, this is a time-consuming task in a model typically consisting of a few millions of sampled points. Instead, the approach of {\tt shapemol} consists of relying on a set of pre-generated tables of $k_\nu$ and $j_\nu$ as functions of $n$ and $T$, each table corresponding to a species (\doce\ and \trece\ in its present state), a value of $\epsilon$, and a value of the $\frac{r}{V} X$ product. The values of $k_\nu$ and $j_\nu$ in each of these tables have been previously computed for each $n$ and $T$ by iteratively solving the population level equations. 

\begin{landscape}
\begin{figure}[!]
%\center
\resizebox{25cm}{!}{\includegraphics{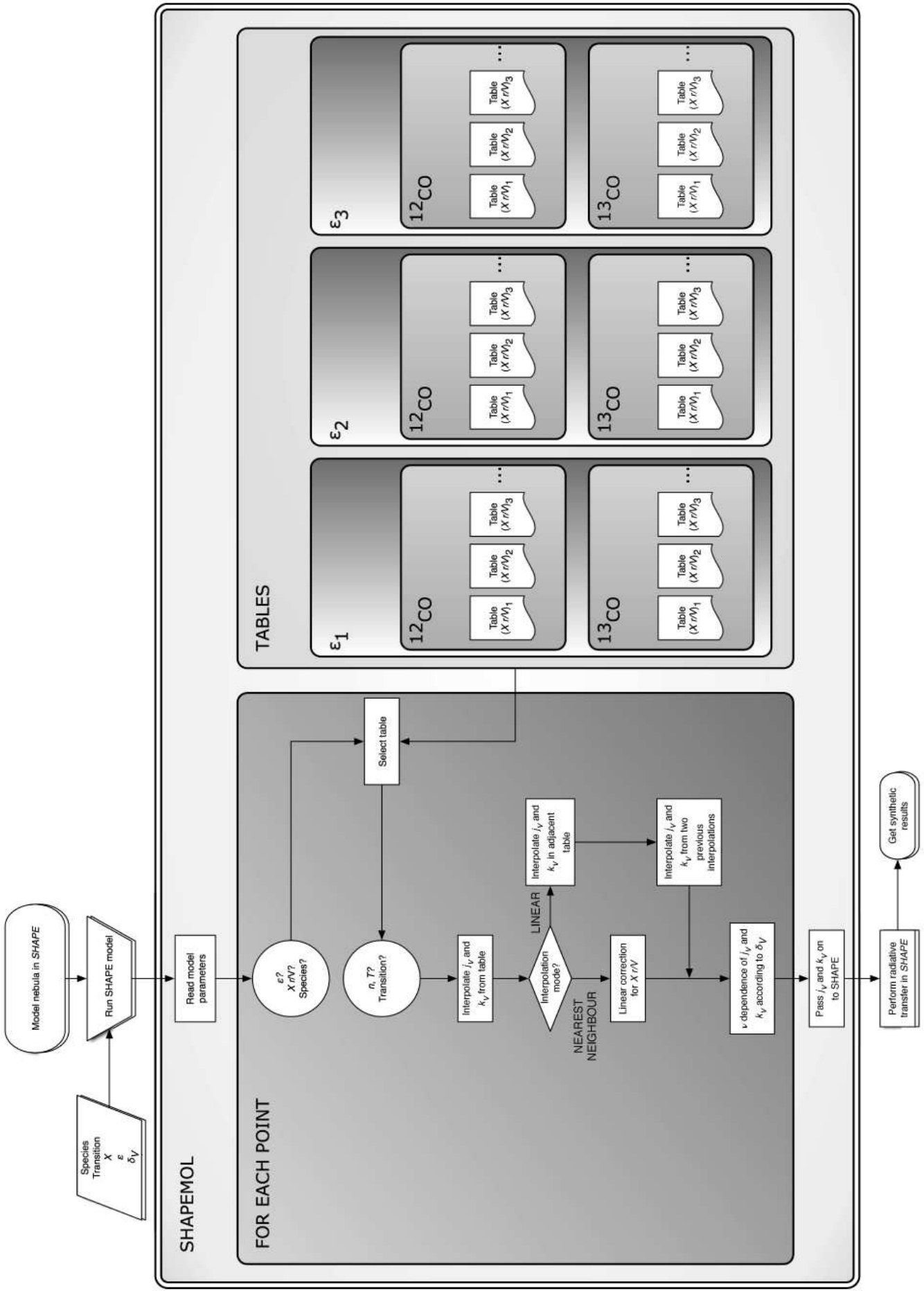}}
\caption{Workflow of {\tt shapemol}.}
\label{FF}
\end{figure}
\end{landscape}

The flow diagram of {\tt shapemol} is shown in Fig.~\ref{FF}. The user selects the desired species and transition in the {\tt SHAPE} interface, together with abundance $X$, $\epsilon$, and $\delta_{\mathrm{V}}$ (the characteristic microturbulence) values of each structure of the model (lobes, polar jets, etc.). The code then samples the positions, velocities, densities, and temperatures of the whole model ---with a spatial resolution chosen by the user---, and computes the product $\frac{r}{V} X$ for each sampled position. Then, it selects the table with the closest value of $\epsilon$ and $\frac{r}{V} X$ for the given species and transition. Once a table has been selected, {\tt shapemol} computes the $k_\nu$ and $j_\nu$ for each sampled position by linear interpolation between the values for the two adjacent tabulated values of $n$ and $T$. The steps in $n$ and $T$ were specifically chosen to be small enough so as to guarantee that a first order interpolation is a good approximation between two consecutive values (see\ref{uncertainty}). Finally, given the roughly linear dependence of $k_\nu$ and $j_\nu$ on the product $\frac{r}{V} X$, the software scales the computed absorption and emission coefficients according to the ratio of the sampled position value of $\frac{r}{V} X$ to that of the selected table; to avoid significant errors, calculations were performed for many values of this parameter.

By default, {\tt shapemol} computes $k_\nu$ and $j_\nu$ by interpolation and applies a linear correction on the coefficients based on the ratio of the sampled position value of $\frac{r}{V} X$ to that of the selected table. If, however, the option ``Linear Interpolation'' is set, {\tt shapemol} selects the two tables whose $\frac{r}{V} X$ is closest to that of the cell, linearly interpolates the values of $k_\nu$ and $j_\nu$ in both tables in order to have the coefficients at those two $\frac{r}{V} X$, and uses those to linearly interpolate the final $k_\nu$ and $j_\nu$ at the cell's $\frac{r}{V} X$. This procedure is slightly more accurate, since no assumptions are made on the dependence of $k_\nu$ and $j_\nu$ on $\frac{r}{V} X$, although it is slightly slower.

In its present state, {\tt shapemol} is able to compute the $J$=1$-$0, 2$-$1, 3$-$2,... up to the 17$-$16 transitions of $^{12}$CO and $^{13}$CO. There are 199 values of $T$ and 21 values of $n$ in the available pre-generated tables, ranging from 10 to 1000 K in steps of 5 K for the temperature, and from 10$^2$ cm$^{-3}$ to 10$^7$ cm$^{-3}$ in multiplicative factors of $\sqrt[4]{10}$ for the density. {\tt shapemol} has been successfully used to study the excitation conditions of the molecular gas of the young planetary nebula NGC~7027, which was found to show a relatively complex physical conditions and structure (\citealp{santander12}).

%where the synthetic profiles have been matched against data from Herschel/HIFI and the IRAM 30-m radiotelescope. The molecular region of this nebula was described as four nested, mildly bipolar shells plus a pair of high-velocity polar blobs. Each of these shells is characterised by single values of the density $n$, temperature $T$, abundance $X$ and velocity $V$. The excitation conditions of the nebula change across shells: the inner, denser ($n\sim$1.5$\times$10$^5$~cm$^{-3}$) and hotter ($T\sim$400~K) shell expands faster and shows a jump in density and temperature, indicative of a shock front moving throughout the system, while the outermost shell is significatively slower, tenuous ($n\sim$5$\times$10$^3$~cm$^{-3}$) and colder ($T\sim25$~K) .

\subsection{Other molecular species}
%\subsection{Future plans}

Future plans include the addition of more molecules, such as C$^{18}$O, CS, and SiO, whenever the effect of vibrational excitation is low enough to warrant results as accurate as those presented here for CO (see Appendix A). In any case, the implementation of {\tt shapemol} allows the inclusion of user-supplied sets of tables containing the absorption and emission coefficients of transitions of other molecular species. When a set of tables in a specific format has been placed in the user's computer and the index text file has been properly modified, {\tt shapemol} will be able to work with the supplied data without the need for compiling a new version of {\tt SHAPE}. For questions or details on the format, contact the first author of this work. 

For simple molecules and when the IR source can be assumed to be small, located in the centre of the nebula, and emitting isotropically, the effect of the IR vibrational excitation could be implemented, at a cost of a considerable increase in the volume of the tables (of the order of at least a factor 20) and a detailed study of the applicability of the treatment itself.

\subsection{Interaction with {\tt SHAPE}}

This is a brief, conceptual description of the workflow when using {\tt SHAPE} together with {\tt shapemol}. 

\begin{enumerate}

\item We model a nebula with several distinct outflows or structures using the standard tools in {\tt SHAPE}. The nebula must have defined values of the density, temperature, and velocity modifiers. Next, we create the desired molecular species ---$^{12}$CO or $^{13}$CO--- and transition for each structure of the nebula, and assign values for the molecular abundance, $\epsilon$, and micro-turbulence, $\delta_{\mathrm{V}}$.

\item When the model is run, the nebula is then sampled, point by point, with a resolution chosen by the user. Each of these points is characterised by an identification code (unique for each structure), a position, a velocity, a density, and a temperature. Only points that are actually populated are processed explicitly, this saves memory and processing time.  {\tt shapemol} then computes the absorption and emission coefficients using the LVG approximation-based tables.

\item {\tt SHAPE} reads the output by {\tt shapemol} and performs radiative transfer solving for the whole nebula, taking into account the emission from the cosmic microwave background, which is subtracted afterwards in order to compute the final line profile. The result is a data-cube consisting of a stack of images of the nebula, one per frequency resolution element (i.e. spectral channel). This data-cube is then convolved, channel by channel, with the telescope beam, which is simulated by a Gaussian with a Full Width at Half Maximum (FWHM) equal to the telescope Half-Power Beam Width (HPBW). The profile from the grid pixel corresponding to the telescope pointing is taken, and its intensity is divided by the projected area of the grid pixel on the plane of the sky. A final profile is then shown, in units of W~s$^{-1}$~Hz$^{-1}$~m$^{-2}$~sr$^{-1}$ or main beam temperature ($T_\mathrm{mb}$), allowing for direct comparison with observations.

\end{enumerate}

\section{Sources of uncertainty}\label{uncertainty}

\subsection{Radiative transfer solving in {\tt SHAPE}}

Radiative transfer solving requires finding the intensity, for each frequency, that matches the following equation at each location of the target (i.e. a nebula) in the plane of the sky:
\begin{equation}
I_\nu(s) ~=~ I_\nu(s_0) e^{-\tau(s)} ~+~ \int_{s_0}^s S_\nu e^{[\tau(s)-\tau(x)]} d\tau(x) ~~,
\end{equation}
where $s_0$ is the position of the edge of the target located farthest away from the observer, $s$ is the position measured from $s_0$ along the line of sight towards the observer, $\tau$ is the optical depth, with its differential $d\tau(x)$ and $x$ the integration variable, also running along the line of sight towards the observer, $I$ is the intensity emitted per unit of time, frequency, and solid angle, and $S$ the source function, which is the ratio of the emission to the absorption coefficient: $S=\frac{j_\nu}{k_\nu}$. In general, $S$ varies along $d\tau(x)$ and thus the solution to the equation is often complex, and can seldom be expressed in analytical form.

The approach that {\tt SHAPE} uses consists of sampling the nebula point by point according to the resolution chosen by the user. Provided that we know the absorption and emission coefficients (and therefore the source function) at each of these points, solving the radiative transfer equation implies accumulating the effects of the emission and absorption coefficients at each point, from i=0 (i.e. the points located farthest away from the observer, along the line of sight) to the last point in the system (i.e. the closest one to the observer). At each step, the emission coefficient contributes to increasing the accumulated intensity, while the absorption coefficient does the opposite (unless it is negative, in which case we have a maser).

This method provides a good approximation to the real case, provided that the number of points (i.e. the spatial resolution of the model nebula) is large enough. However, the accumulated sum of the effects of emission and subtraction can result in computational rounding errors. To test to which extent this occurred within {\tt SHAPE}, we modelled a simple nebula for which an analytical solution exists under several assumptions.

The common ground for each of these models was a spherical nebula with radius $r=1.2 \times 10^{16}$~cm located 100~pc away (projecting to a radius of 8 arcsec on the plane of the sky) and filled with gas with homogeneous temperature $T=100$~K and density $n=1.5 \times 10^5$~cm$^{-3}$, and micro-turbulence $\delta_V=2.5$~km s$^{-1}$. Under the assumption of homogeneity, it can be shown that in the optically thick case, the emitted intensity is given by the source function:
\begin{equation}
I_{thick} (W\ Hz^{-1}\ m^{-2}\ sr^{-1}) ~=~ \frac{2 h \nu^3}{c^2} \frac{1}{e^{\frac{h \nu}{K_B T_\mathrm{ex}}}-1} ~~,
\end{equation}
where $I$ is the emitted intensity per unit of time, frequency, surface, and solid angle, $T_\mathrm{ex}$ the excitation temperature, $h$ is the Planck constant and $k_B$ is the Boltzmann constant. In the general case, the intensity is decreased from that in the thick case by a factor that depends on the value of $\tau$ as
\begin{equation}
I ~=~ I_{thick} ~ (1-e^{-\tau})~~,
\end{equation}

For the $^{12}$CO $J$=6$-$5 transition and the given parameters, $I_{thick} = 1.24 \times 10^{-14} W\ Hz^{-1}\ m^{-2}\ sr^{-1}$.

In all our tests, we used a grid consisting of 64$^3$ image pixels that covered a cube with a side length $3 \times 10^{16}$~cm inside of which the spherical nebula was located. We always considered a microturbulence value of $\delta_V = 2.5$~km~s$^{-1}$ and performed the calculations in {\tt SHAPE} using 200 spectral bands for adequately sampling the given spectral profile.

\subsubsection{Static case}

We first considered a nebula without expansion and in thermal equilibrium, so the level population is adequately described within the thermalised-case hypothesis and the absorption and emission coefficients $k_\nu$ and $j_\nu$ are trivial to compute. It must be stressed that we did not make use of the LVG  approximation in these tests (i.e. the coefficients were not computed by {\tt shapemol}), since we are hereby checking the accuracy of radiative transfer itself within {\tt SHAPE}. In this case, the optical depth is simply $\tau = k_\nu L$, where $L$ is the total length of the nebula along the line considered (i.e. in a spherical nebula, $L=2 \times r$ if we look at the centre, and progressively smaller as we consider points nearer the edge).

\begin{figure}[!h]
\center
\resizebox{8cm}{!}{\includegraphics{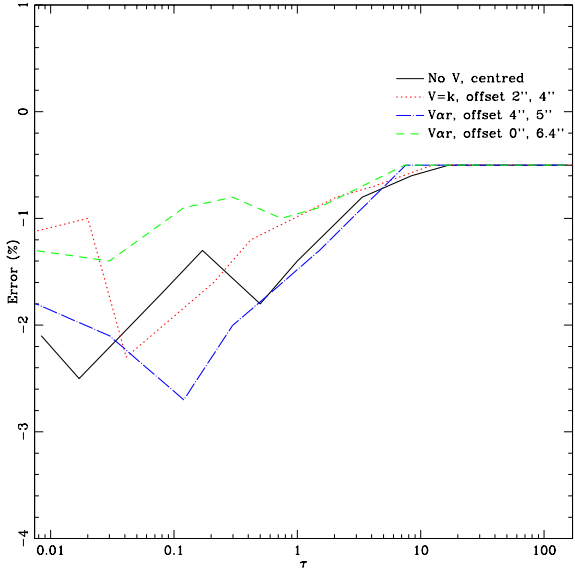}}
\caption{Error behaviour in {\tt SHAPE} radiative transfer as a function of $\tau$ for several cases discussed in the text. The nebula is always a filled sphere with radius 8 arcsec. In the cases with offset, the first number corresponds to a pointing offset from the centre of the spherical nebula on the plane of the sky, while the second number refers to the offset towards the observer (i.e. the intensity is not measured at the profile peak at zero velocity, but at a certain point in the line wings, at a velocity corresponding to the emission in the examined region).}
\label{FX}
\end{figure}

To test the model, we integrated the emission over a square box with side $3 \times 10^{16}$~cm (6 grid pixels, or 1.875 arcsec at the adopted distance) centred on the nebula and compared the values with the theoretical prediction mentioned above for several values of the optical depth (achieved by varying the abundance $X$ over a wide range). The intensity of our {\tt SHAPE}+{\tt shapemol} model almost perfectly matched the analytical value at large $\tau$ (i.e. the optically thick case), $9.8 \times 10^{13} W\ Hz^{-1}\ sr^{-1}$ according to Eq.~3. Errors slowly increase (in absolute value) with decreasing optical depth, as shown in Fig.~\ref{FX} (black thick line), where the error is defined as Error (\%)= 100 $\times \frac{I_S-I_P}{\frac{I_S+I_P}{2}}$, where $I_S$ is the intensity as computed via radiative transfer by {\tt SHAPE}, and $I_P$ the predicted, analytically obtained intensity.

At large $\tau$, the radiative transfer very well matches the analytical prediction, with the error limited to -0.5\% (negative meaning that the analytically calculated intensity is higher than that obtained via radiative transfer). This systematic discrepancy is to be expected, given the difference in the method used for computing the intensities: while the analytical prediction is based on the source function, the absorption and emission coefficients we have used in the radiative transfer come from population level computations based on the theoretical coefficients under the assumption of the thermalised case.

The behaviour of the error changes with decreasing $\tau$, growing in absolute value until it stabilises at around -2\% with small $\tau$. Part of this uncertainty can be explained by the fact that the $\tau$ value used in the analytical prediction was the one corresponding to the centre of the box, while the curvature of the nebula makes the product $k L$ slightly smaller as one considers points offset from the centre. This slight overestimate of the analytical $\tau$ results in a higher opacity and therefore a higher intensity. In this case, with the square box centred on the nebula, this phenomenon can explain approximately one third of the total uncertainty. We can consider the rest as a combination of software rounding errors and the limited spatial resolution (i.e. discretization).

\subsubsection{Constant velocity case}

We next tested the case of a filled, spherical nebula expanding at a constant velocity of $V$=25~km~s$^{-1}$. With a ratio $\frac{V}{\delta_V}$=10 we are well within the assumption of the LVG approximation and can express the optical depth $\tau$ at a given position in the nebula as
\begin{equation}
\tau_{LVG} ~=~ k ~ \sqrt{\pi} \delta_{\nu} \frac{r / (\nu \frac{V_r}{c})}{1+\mu^2 (\epsilon-1)} ~~,
\end{equation}
where $k$ is the normalized absorption coefficient, $V_r$ is the radial component of the velocity, $r$ is the linear distance from the centre of the nebula to the point considered, and $\mu = \cos{\alpha}$, with $\alpha$ the angle between the velocity vector and the line of sight.

In this case, we integrated the emission over a box the same size as in the static case, but centred $3 \times 10^{15}$~cm away from the centre of the spherical nebula (2 arcsec on the plane of the sky), and considered the emitted intensity not at the centre of the spectral profile (i.e. corresponding to V$_r$=0 ~km~s$^{-1}$) like in the previous case, but at a certain point in the profile wings. In particular, we examined the region located $6 \times 10^{15}$~cm towards the observer, which, together with the offset of the box on the plane of the sky, corresponds to the emission at -22.4~km~$s^{-1}$ of the integrated spectral profile.

The error (see Fig.~\ref{FX}, red dotted line) is the same as in the static case at large $\tau$ (i.e. optically thick nebula), and is due to the same reason. The absolute value of the error slowly grows as $\tau$ decreases, reaching values between -1\% and -2.3\%. The reasons for this uncertainty are the same as in the previous, static case.

\subsubsection{Linear velocity case}

Most PNe and PPNe have an approximately ballistic expansion (i.e. under the effect of no forces) after a brief period of shaping; they are said to follow a linear velocity law, with the velocity of each gas particle proportional to the distance of the particle to the star. To reflect this behaviour, we tested the case of a filled, spherical nebula similar to the previous case except for its velocity, with $V \propto r$. The velocity was 0 at the location of the central star, and 25~km~s$^{-1}$ at the edge of the nebula. The LVG approximation can be used here, with the advantage that the equation for $\tau_{LVG}$ becomes simpler since $\epsilon$=1 and the dependence on $\mu$ is removed.

As in the previous case, we considered the emitted intensity at a location away from the centre of the nebula both on the plane of the sky and towards the observer. We examined two such locations, the former centred $6 \times 10^{15}$~cm away from the centre of the nebula (4 arcsec on the plane of the sky) and $7.5 \times 10^{15}$~cm towards the observer (corresponding to the emission at -15.6~km~s$^{-1}$); the latter aligned with the centre of the nebula but offset $9.6 \times 10^{15}$~cm towards the observer (corresponding emission at -20~km~s$^{-1}$).

The errors behave, in both cases, in a similar way to that of the static case: The error is fixed at -0.5\% at large $\tau$, and moderately grows with decreasing $\tau$ until reaching stable values around -2\% in the optically thin regime (see Fig.~\ref{FX}, blue dotted-dashed and green dashed lines). These errors are attributable to the same factors as in the previously discussed static case.

\subsection{Linear interpolation}\label{interpolation}

{\tt shapemol} relies on pre-generated tables of the absorption and emission coefficients $k_\nu$ and $j_\nu$ for a broad range of abundances, $\frac{r}{V}$ ratios, temperatures, and densities. Once a table has been selected based on the product $\frac{r}{V} X$, the given coefficient is linearly interpolated from the adjacent tabulated values in temperature and density. Given the smooth but non-linear distribution of the values and the discretization applied (199 values of the temperature between 10 and 1000~K, 21 values of the density between 10$^2$ and 10$^7$~cm$^{-3}$), this introduces an additional source of uncertainty.

To test the relative importance of this source of uncertainty, we considered the difference between the linearly interpolated value and the ``real'' value at a given point as the second-order term of the Taylor-series development of the desired coefficient $k_\nu$ or $j_\nu$ as a function of either the density or the temperature, with the other one remaining fixed. We estimated this term by computing the second derivative of the function in the neighbouring points of the table for every point. The associated error was given as a percentage of the value of the $k$ or $j$ coefficient at that point. However, since such a procedure would lead to divisions by almost zero wherever the coefficient is close to zero, and thus to unrealistic errors, we only took into account values within the top 80\% of the highest value of the coefficient $k_\nu$ or $j_\nu$ for the given function.

The resulting errors for the $k_\nu$ and $j_\nu$ coefficients are below 1\% for the vast majority of the tables, with some peculiarities worth noting: the error is slightly larger at low-$J$ transitions (i.e. $J$=1$-$0 to 4$-$3) and very low temperatures, in particular 15~K. In that case, the error reaches 2.8\% for both $k_\nu$ and $j_\nu$, except in the 1$-$0 transition at 15~K, where the $k$ coefficient shows a 7\% error.

\section{Limitations}\label{limitations}

This section briefly describes the limitations of {\tt shapemol}, which the user should be aware of.

\subsection{Very faint emission}

{\tt shapemol} is not meant for studying very faint emission from extremely sub-excited, high-$J$ transitions. For instance, the theoretical emission coefficient $j_\nu$ for the \doce\ $J$=16$-$15 transition at a temperature of 15~K (at which these levels have a very low population) is  $\sim$10$^{10}$ larger than at 10~K. In {\tt shapemol},  we have set every value of $j_\nu$ and $k_\nu$ below a certain very restrictive threshold to 0 to avoid computational noise. In particular, the coefficients are set to 0 wherever the energy (in K units) of the levels involved is $\gtrsim$28 times higher than the corresponding equivalent rotational temperature describing the level population. In practice, however, cutting off the emission and absorption from high-$J$ transitions at very low temperatures is unlikely to become a problem, since observational noise is usually much higher than the emission one would get from such a sub-excited transition, and no such emission is detected in actual cases. In most cases, moreover, the contribution from other nebular components that are not this sub-excited will completely dominate the total emission. The user should nevertheless be aware that for very faint emission (i.e. at a level below $\sim$5\% of the highest value of the coefficient $j_\nu$ or $k_\nu$ in a given table and transition), errors in the linear interpolation discussed in Sect. \ref{interpolation} will be significantly larger than stated there. In particular, errors in the interpolation will be large when crossing the jump from a zero value to the first non-zero value. 

At these faint levels of emission, the contribution of the IR continuum from the central star or surrounding dust (i.e. the effect of vibrational transitions, see Appendix A), which is not accounted for by {\tt shapemol}, could be significant.

\subsection{Abuse of LVG conditions}

The LVG approximation adequately describes the physical excitation conditions in most cases, even when the velocity is similar to the microturbulence value, $\delta_\mathrm{V}$ (see details in \citealp{bujarrabal13}). It can be applied to almost all PNe and PPNe, except in the most peculiar objects. However, in some cases, this approximation can give no more than a rough, qualitative idea of the excitation conditions of the target. The reason is that the emission and absorption coefficients, $j_\nu$ and $k_\nu$, might be overestimated or underestimated when the LVG coherence zone (the region where particles might have similar velocities and thus are prone to photon absorption and interaction) is much larger or smaller than in the actual nebula.

%Also, it does not work for zero velocities.

Consider, for example, a shell expanding at a very low velocity (i.e. similar to $\delta_{\mathrm{V}}$). For any considered location in the shell, the corresponding LVG coherence zone will be a large, imaginary sphere centred on that location. The LVG approximation will consider photons coming from every point inside that sphere in the absorption and emission coefficient calculation (i.e. the effect usually called photon-trapping). However, if the shell is thin and the velocity is low enough, the LVG coherence zone sphere will be larger than the width of the shell, thus reaching farther towards empty space. {\tt shapemol}, using the LVG approximation in the level population calculations, will then falsely consider absorption and emission from photons coming from an empty region. In other words, photon-trapping will be overestimated, and so $k_\nu$ and $j_\nu$ will be, specially for high-$J$ transitions.

Another example worth considering is that of a disk in Keplerian rotation. Here, when $V$ is very low or high (at large or small distances from the star respectively), the LVG coherence zone is too large or small compared to the actual photon-trapping region, and $k_\nu$ and $j_\nu$ will be incorrectly computed. This leads to a result that, although it can give an idea of the excitation conditions in the disk, cannot be considered quantitatively accurate. 

Note, however, that this problem only affects the LVG calculations of $j_\nu$ and $k_\nu$; instead, radiative transfer solving in {\tt SHAPE} remains unaffected and is as precise as in any other case.

\subsection{Masers}

Maser emission will occur whenever the absorption $k_\nu$ coefficient is $<$0. This occurs for rare, specific values of $n$, $T$, and $\frac{r}{V} X$, and mostly in very low-$J$ CO transitions. Each point along the line of sight with such an absorption coefficient will amplify the arriving intensity. Real masers in CO are never very intense (a maser is considered intense when $\tau<-1$) under conditions of collisional excitation, since the pumping of CO masers by collisions is always inefficient. However, {\tt shapemol} could spuriously yield high amplification if many of the model points share those specific values of the parameters and thus many points with negative $k_\nu$ are accumulated along the line of sight. In such a situation, which is in fact extremely improbable, a false maser several orders of magnitude more intense than the true emission will be conspicuous in the synthetic spectral profile. Note, however, that this problem can only occur when the LVG calculations of $j_\nu$ and $k_\nu$ are used for real conditions that would in fact not allow the use of the LVG approximation.

In the current version of {\tt SHAPE}, a warning is displayed at the radiative transfer solving stage whenever $\tau<-1$ in a given line of sight and velocity.

%\subsection{Large deviation in $\frac{r}{V}$}\label{rv_abuse}

%If the deviation of $\frac{r}{V}$ in a given structure is significant, the table used for computing the $k_\nu$ and $j_\nu$ coefficients in that structure will not be adequately representative of the physical conditions in many of the grid cells of the structure. In cases, for example, where a structure is very extended in the radial direction from the star but has a constant or nearly-constant velocity (e.g.\ a polar jet not showing a Hubble-flow),  $\frac{r}{V}$ will be very different near the star and at near the outer edge of the structure. {\tt shapemol} will compute the mean value of  $\frac{r}{V}$ in the structure and use the corresponding table (based on  the resulting $\frac{r}{V} X$) for computing the $k_\nu$ and $j_\nu$ coefficients of the whole structure (and then, if the keyword quick\_XrV is set, apply a linear correction according to that mean value of $\frac{r}{V} X$), thus returning wrong values of the coefficients if the ratio $\frac{r}{V}$ changes significantly across the structure. In general, if the difference in $\frac{r}{V}$ between opposite edges of the structure is larger than $\sim$25\% (the difference in $\frac{r}{V} X$ between consecutive tables), the user should consider splitting the given structure in two or more smaller structures.

\section{Applying shapemol: the case of NGC 6302}\label{science}

\begin{figure*}[h!]
\center
\resizebox{11cm}{!}{\includegraphics{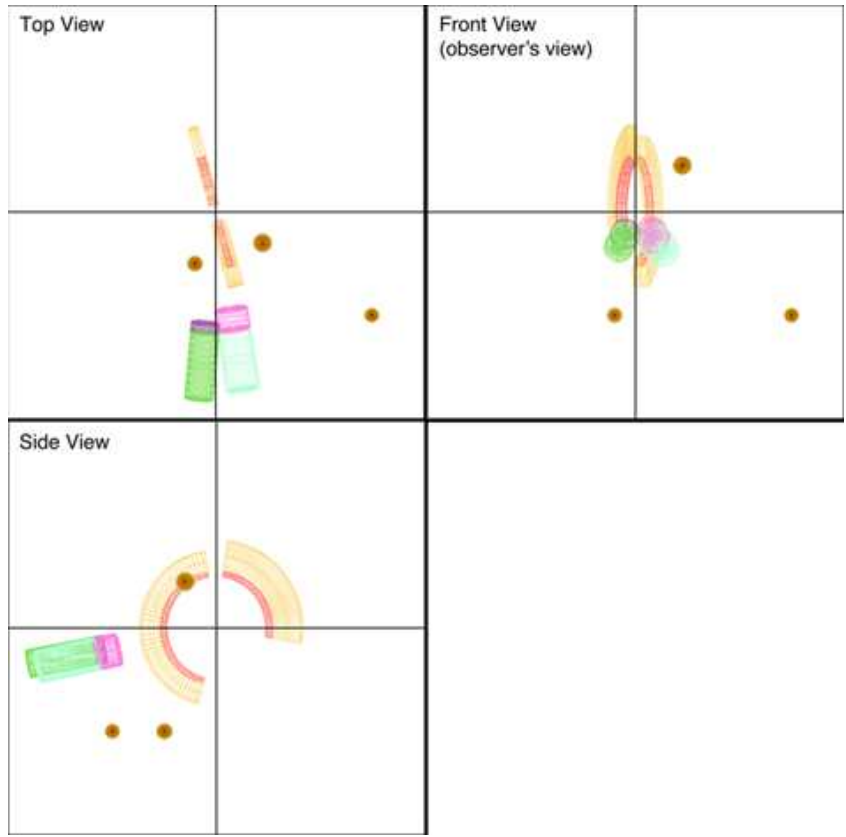}}
\caption{3-D mesh view of the model of the molecular envelope of NGC 6302. The size of each frame is 40$\times$40 arcsec$^2$, and its label corresponds to the usual convention within {\tt SHAPE}: {\it Front View}, {\it Top View} and {\it Side View} correspond to the view from Earth (north up, east to the left), from the direction defined by north in the plane of the sky, and from the direction defined by west in the plane of the sky (so the observer is on the left).}
\label{FNGC6302_1}
\end{figure*}

\begin{figure*}[h!]
\center
\resizebox{16cm}{!}{\includegraphics{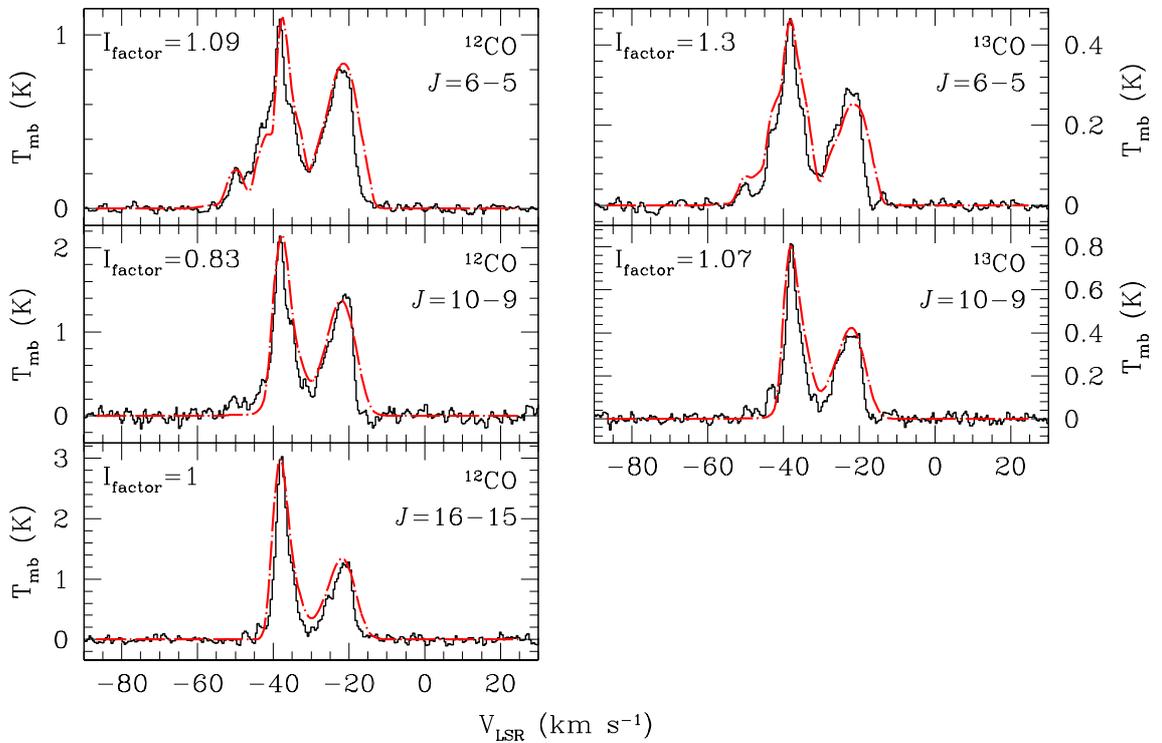}}
\caption{Resulting synthetic spectra (red, dotted-dashed line) and observations (black histogram) for the $^{12}$CO and $^{13}$CO transitions detected in NGC~6302 with HIFI. $I_\mathrm{factor}$ refers to the intensity factor applied to the model to account for the uncertainties in the radiative transfer solving and in calibrating the observations.}
\label{FNGC6302_2}
\end{figure*}

\begin{figure*}[h!]
\center
\resizebox{12cm}{!}{\includegraphics{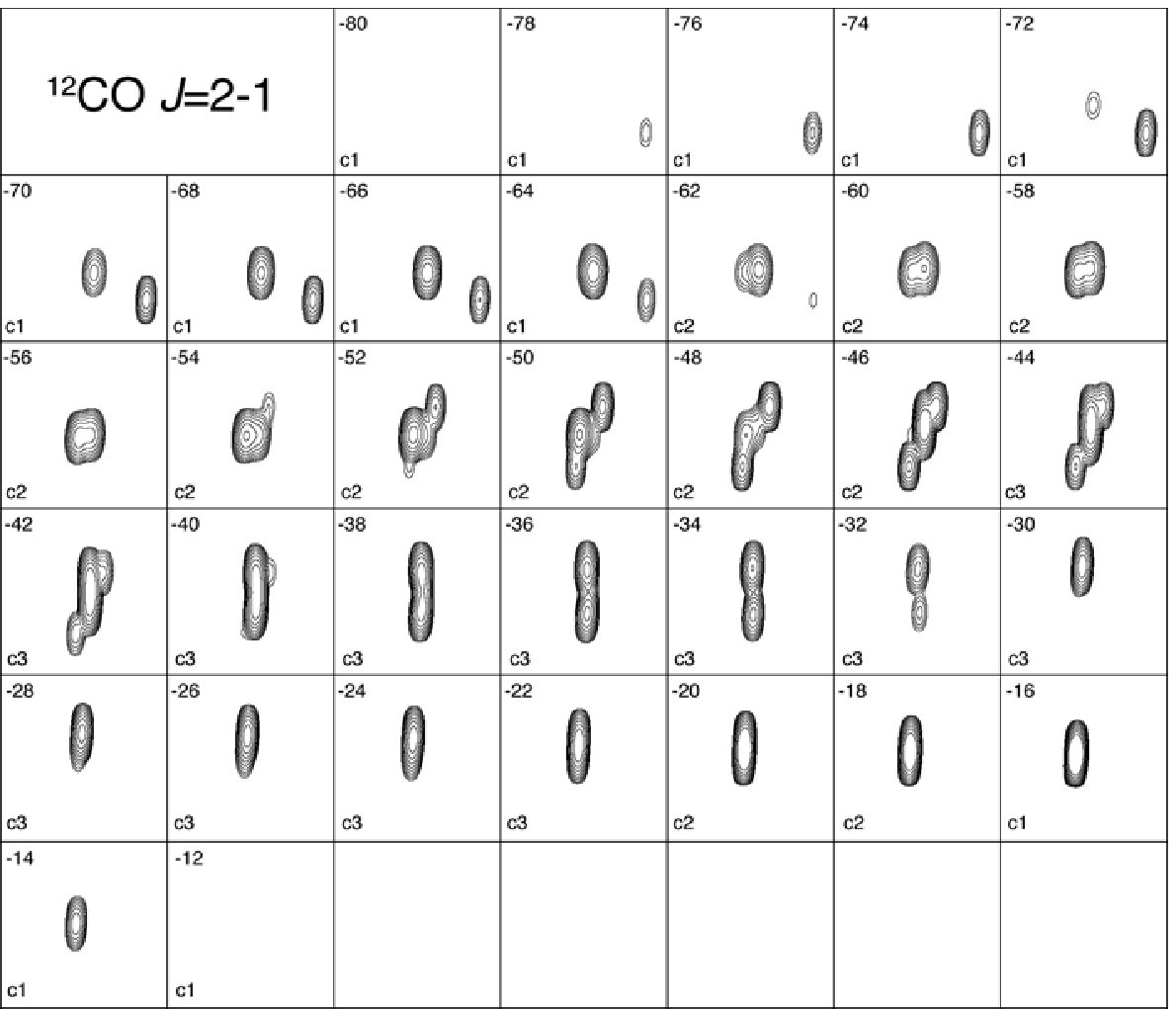}}
\caption{Resulting synthetic interferometric map of NGC 6302 in the \doce\ \jdu\ transition. The size of each frame is 20$\times$20 arcsec$^2$. The LSR velocity is indicated in the upper left corner of each frame. The contour levels have been adapted for approximate visual comparison with the observations displayed in Fig.~1 in \cite{dinhvtrung08};  emission is represented in a logarithmic scale, with the following maxima (10$^\mathrm{th}$ contour) depending on the label: 2.4 Jy~beam$^-1$ for `c1' channels, 6.4 Jy~beam$^-1$ for `c2' channels, and 11 Jy~beam$^-1$ for `c3' channels.}
\label{FNGC6302_3}
\end{figure*}

\begin{figure*}[h!]
\center
\resizebox{12cm}{!}{\includegraphics{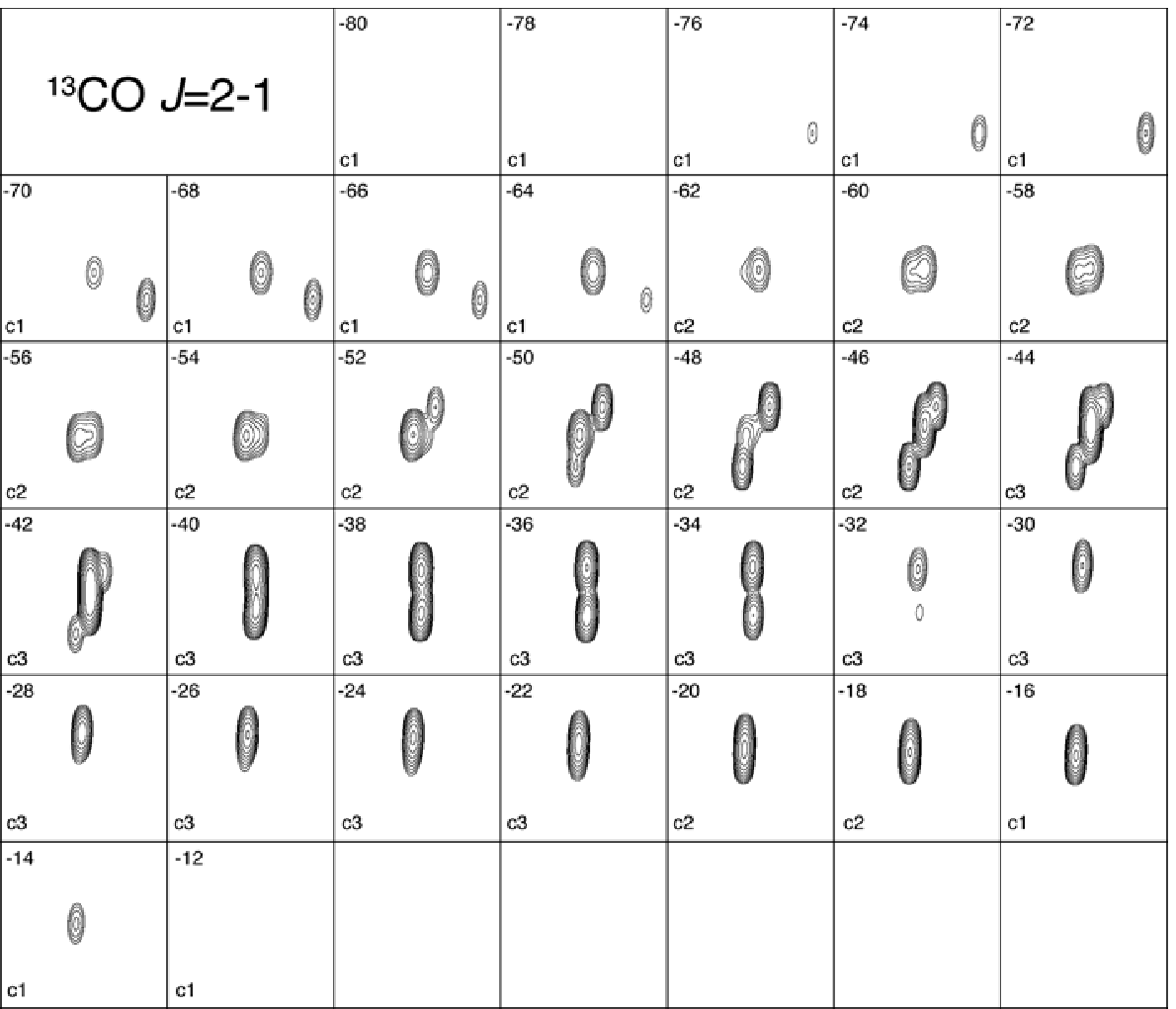}}
\caption{Resulting synthetic interferometric map of NGC 6302 in the \doce\ \jdu\ transition. The size of each frame is 20$\times$20 arcsec$^2$. The LSR velocity is indicated in the upper left corner of each frame. The contour levels have been adapted for approximate visual comparison with the observations displayed in Fig.~2 in \cite{dinhvtrung08};  emission is represented in a logarithmic scale, with the following maxima (10$^\mathrm{th}$ contour) depending on the label: 2.7 Jy~beam$^-1$ for `c1' channels, 5.4 Jy~beam$^-1$ for `c2' channels, and 7.5 Jy~beam$^-1$ for `c3' channels.}
\label{FNGC6302_4}
\end{figure*}

\begin{table*}\renewcommand{\arraystretch}{1.3}
  \begin{center}
  \caption{Best-fit model parameters for the molecular ring components of NGC 6302. The sizes correspond to a distance to the nebula of 1.17~kpc. The uncertainty range is included for every parameter except for those concerning the geometry.}
  \label{T11}
  \begin{tabular}{|l|c|c|c|c|c|c|c|c|c|c|c|}\hline 
{\bf Structure} & $r_\mathrm{inner}$ &  $r_\mathrm{outer}$ & $h$  &  $\varphi_1$ &  $\varphi_2$ &  P.A. & $i$ &  $\delta_\mathrm{V}$ & $n$ & $T$ &  $V_\mathrm{exp}$ \\   
 & (10$^{16}$~cm) & (10$^{16}$~cm) & (10$^{16}$~cm)  &  &  & &  &  (km~s$^{-1}$) & (cm$^{-3})$ & (K)  &  (km~s$^{-1}$) \\   
\hline
\hline
\multicolumn{11}{|l|}{{\bf Inner ring}} & constant value \\
\hline
Blue inner ring & 8.8 & 9.7 & 1.1 & -165$^\mathrm{o}$  & -10$^\mathrm{o}$  & 270$^\mathrm{o}$ & -75$^\mathrm{o}$ & 2 & 2.3$^{+0.7}_{-0.3}\times$10$^{5}$ & 260$^{+35}_{-20}$ & 8.5$^{+0.5}_{-0.5}$  \\
Red inner ring & 8.8 & 9.7 & 1.8 & 8$^\mathrm{o}$  & 100$^\mathrm{o}$  & 270$^\mathrm{o}$ & -75$^\mathrm{o}$ & 2 & 2.4$^{+0.4}_{-0.2}\times$10$^{5}$ & 300$^{+30}_{-30}$ & 11$^{+2}_{-2}$  \\
\hline
\hline
\hline
\multicolumn{11}{|l|}{{\bf Outer ring}} & linear $V$ max.\\
\hline
Blue outer ring & 9.7 & 13.2 & 2.6 & -165$^\mathrm{o}$  & -10$^\mathrm{o}$  & 270$^\mathrm{o}$ & -75$^\mathrm{o}$ & 1 &1.6$^{+1.4}_{-1.1}\times$10$^{5}$ & 40$^{+10}_{-5}$ & 14$^{+1}_{-1}\ ^\dagger$  \\
Red outer ring & 9.7 & 14.9 & 1.8 & 8$^\mathrm{o}$  & 100$^\mathrm{o}$  & 270$^\mathrm{o}$ & -75$^\mathrm{o}$ & 1 &4$^{+4}_{-2}\times$10$^{4}$ & 60$^{+20}_{-25}$ & 17$^{+1}_{-1}\ ^\dagger$  \\
\hline
\hline
  \end{tabular}
 \end{center}
\vspace{1mm}
 \scriptsize{
  {\it Parameters:}  \\
   $r_\mathrm{inner}$ and $r_\mathrm{outer}$ are the inner and outer radii of the ring, and $h$ is its thickness. \\
   $\varphi_1$ and $\varphi_2$ are the lowest and highest values of the angular span of the ring, as defined within {\tt SHAPE}. \\
   P.A. and $i$ are the position angle and inclination of the symmetry axis of the structure, as defined within the {\tt SHAPE} modifier `PA/Inc Rotation'. \\
   $\delta_\mathrm{V}$, $n$, and $T$ are  the microturbulence velocity, the density, and the temperature, respectively. \\  
  $V_\mathrm{exp}$ is the expansion velocity. In the case of the inner ring, it is a constant value. The value of the velocity shown for the outer ring is the velocity of the farthest edge from the star, with the velocity following a linear, ballistic expansion pattern. \\
}
\end{table*}

\begin{table*}\renewcommand{\arraystretch}{1.3}
  \begin{center}
  \caption{Best-fit model parameters for the blob components of NGC 6302. The sizes correspond to a distance to the nebula of 1.17~kpc. The uncertainty range is included for every parameter except for those concerning the geometry.}
  \label{T22}
  \begin{tabular}{|l|c|c|c|c|c|c|c|c|c|}\hline 
{\bf Structure} & $r$ &  $x_\mathrm{offset}$ & $y_\mathrm{offset}$ & $z_\mathrm{offset}$ &  $\delta_\mathrm{V}$ &  $n$ & $T$ & $V_\mathrm{exp}$ \\   
 & (10$^{16}$~cm)  &  (10$^{16}$~cm)  &  (10$^{16}$~cm)  &  (10$^{16}$~cm) & (km~s$^{-1}$) & (cm$^{-3})$ &  (K)  & (km~s$^{-1}$) \\   
\hline
\hline
Blob N & 1.6 & 7.9 & 7.9 & -5.3  &  2 & 5$^{+4}_{-2}\times$10$^{5}$ & 40$^{+20}_{-10}$ & 35$^{+3}_{-2}$  \\
Blob S & 1.3 & -3.5 & -17.6 & -8.8  &  2 & 5$^{+3}_{-2}\times$10$^{5}$ & 40$^{+20}_{-10}$ & 32$^{+2}_{-2}$  \\
Blob SE & 1.2 & 26.3 & -17.6 & -17.6  &  4 &5$^{+3}_{-2}\times$10$^{4}$ & 20$^{+20}_{-5}$ & 78$^{+3}_{-3}$  \\
\hline
\hline
  \end{tabular}
 \end{center}
\vspace{1mm}
 \scriptsize{
  {\it Parameters:}  \\
   $r$ is the radius of the blob. \\
   $x$,$y$,$z$ are the offsets of the blob with respect to the central star, defined with $x$ increasing towards West, $y$ increasing towards North, and $z$ increasing along the observer-object line. \\
   $\delta_\mathrm{V}$, $n$, and $T$ are the microturbulence velocity, the density, and the temperature, respectively. \\  
  $V_\mathrm{exp}$ is a single, constant value representing the expansion velocity.
  }
\end{table*}

\begin{table*}\renewcommand{\arraystretch}{1.3}
  \begin{center}
  \caption{Best-fit model parameters for the finger components of NGC 6302. The sizes correspond to a distance to the nebula of 1.17~kpc. The uncertainty range is included for every parameter except for those concerning the geometry.}
  \label{T33}
  \begin{tabular}{|l|c|c|c|c|c|c|c|c|c|c|}\hline 
{\bf Structure} & $r$ &   $h$   &  offset &  P.A. &  $i$ &  $\delta_\mathrm{V}$ & $n$ &  $T$ & $V_\mathrm{exp}$ \\   
 & (10$^{16}$~cm)  &  (10$^{16}$~cm)  &  (10$^{16}$~cm)  &  & & (km~s$^{-1}$) &  (cm$^{-3})$ &  (K)  &  (km~s$^{-1}$) \\   
\hline
\hline
\multicolumn{9}{|l|}{{\bf Finger inner sections}} & constant value \\
\hline
Finger E inner & 2.5 & 1.1 & 20.2 & 155$^\mathrm{o}$  &  192$^\mathrm{o}$ & 2 & 3$^{+1}_{-1}\times$10$^{4}$ & 80$^{+45}_{-20}$ & 20$^{+1}_{-1}$  \\
Finger W inner & 2.8 & 3.5 & 18.4 & 215$^\mathrm{o}$  &  195$^\mathrm{o}$ & 4 & 8$^{+4}_{-2}\times$10$^{3}$ & 70$^{+30}_{-30}$ & 26$^{+2}_{-2}$  \\
\hline
\hline
\multicolumn{9}{|l|}{{\bf Finger outer sections}} & linear $V$ max.\\
\hline
Finger E outer & 2.5 & 11.4 & 26.8 & 155$^\mathrm{o}$  &  192$^\mathrm{o}$ & 2 & 4$^{+2}_{-0.9}\times$10$^{3}$ & 50$^{+40}_{-20}$ & 28$^{+2}_{-2}$  \\
Finger W outer & 2.8 & 10.5 & 25.9 & 215$^\mathrm{o}$  &  195$^\mathrm{o}$ & 2 & 2$^{+0.7}_{-0.3}\times$10$^{3}$ & 50$^{+35}_{-20}$ & 38$^{+2}_{-2}$  \\
\hline
\hline
  \end{tabular}
 \end{center}
\vspace{1mm}
 \scriptsize{
  {\it Parameters:}  \\
   $r$ is the radius of the cylinder defining the finger, and  $h$ its height. \\
   ``offset'' refers to the displacement of the centre of a given structure from the central star, in the direction defined by P.A. and $i$. \\
     P.A. and $i$ are the position angle and inclination of the symmetry axis of the structure, as defined within the {\tt SHAPE} modifier `PA/Inc Rotation'. \\
   $\delta_\mathrm{V}$, $n$, and $T$ are the microturbulence velocity, the density, and the temperature, respectively. \\  
  $V_\mathrm{exp}$ is the expansion velocity of the given structure. The velocity of the inner section of each finger is a constant value. The value of the velocity shown for the outer section is the velocity of the farthest edge from the star, with the velocity following a ballistic expansion pattern. 
  }
\end{table*}

\subsection{Motivation}

NGC 6302 is a spectacular young PN. Located 1.17$\pm$0.19~kpc away (\citealp{meaburn08}), it hosts a very hot central star with a T$_\mathrm{eff}$ of at least 150,000 K (\citealp{wright11}). The nebula shows an extreme bipolar morphology in optical images, along with an equatorial dust lane that most likely is associated with a molecular torus or disk (\citealp{matsuura05}). Authors have derived different values of the mass of this molecular structure, which, when scaled to the distance adopted in this work (with mass depending on the distance squared), can be used to establish valid comparisons. In particular, \cite{peretto07} analysed low-excitation CO interferometric observations and found an expanding torus with a mass of $\sim$2.7 M$_\odot$. This is completely at odds with the masses derived by other authors: \cite{gomez89} found $\sim$0.14 M$_\odot$, while \cite{huggins89} found a slightly higher mass of 0.3 M$_\odot$.  \cite{dinhvtrung08} computed a mass of $\sim$0.1 M$_\odot$ through the means of \doce\ and \trece\ \jdu\ interferometric mapping. More recently, \cite{bujarrabal11} presented high-$J$ CO transitions observed with Herschel/HIFI and provided some constraints in density and temperature that were compatible with the findings by \citeauthor{dinhvtrung08}

Careful consideration of the data leads to the conclusion that the molecular component of the nebula must be broken into clumps of dense material, with other regions completely devoid of molecules. In other words, a simple toroidal model with constant values of the density and temperature provides a rudimentary fit to the data, at most. In an effort to better constrain the mass and excitation conditions of the nebula, we have built a model of the molecular component of NGC 6302 and compared it with both the low-$J$ interferometric map by \citeauthor{dinhvtrung08} and the high-$J$ observations presented by \cite{bujarrabal11}. 

The main motivation behind this model is to show the potential of {\tt shapemol}. Because it is integrated within {\tt SHAPE} and because of the 3-D engine the model can be modified and adapted with a few mouse clicks, including its geometry, which can be as complex and detailed as needed. On the other hand, radiative transfer modelling with {\tt shapemol} runs in a few tens of seconds on current standard computers, allowing quick iteration while still providing a reasonable approximation to the actual physical conditions. 

Additionally, this model shows  the complexity of the molecular envelope of NGC 6302 and allows us to determine the physical and chemical properties of its molecular envelope, as well as to provide a full description of the 3-D morphology of the nebula assuming a homologous expansion for the velocity field. We investigate the characteristic densities and temperatures of the nebula and provide an independent, alternate determination of the molecular mass of the nebula.

\subsection{The model}

The model consists of a broken torus with the same orientation as the simple model by \citeauthor{dinhvtrung08} plus a series of fingers and blobs expanding outwards from the central star (see the sketch in Fig.~\ref{FNGC6302_1}). This torus is broken into two main regions, named {\it blue} and {\it red} according to their Doppler shift with respect to the central star. Each of these regions is a piece of a cylindrical ring, each different in size and span, and is itself split into two components: an {\it inner}, thinner, hotter, and much denser component, and an {\it outer}, larger, cooler, and more tenuous one. The model also contains three spherical {\it blobs} with different physical conditions, and two outward-projected cylindrical filaments we have called {\it fingers}, each one split into an inner and an outer section. 

All these components are in fact directly identifiable in the interferometric maps, which show emission from a structure resembling a broken torus with smaller condensations at different positions and velocities (see Fig.~1 in \citealp{dinhvtrung08}). The positions of several of these condensations show a dependence on their velocity. Careful consideration of the high-$J$ profiles from HIFI allows us to associate some of these structures with their profile counterparts, revealing that the torus must have an inner region with a higher temperature and density.

To keep the model as simple as possible while providing a good fit to the data, we have fixed the temperature and density of each structure to a single, representative value. All the structures are expanding away from the star with velocity laws that are either constant values (in structures that are too thin, in the radial direction, to produce an observable difference in the profile) or ballistic velocity patterns (in radially thicker structures). Each structure has its own micro-turbulence velocity value, $\delta_{\mathrm{V}}$. On the other hand, we have set the \doce\ and \trece\  abundances to two unique values for the whole nebula. 

Fig.~\ref{FNGC6302_1} shows a three-dimensional view of the model with all structures visible. We fitted the model in the same iterative fashion as in the modelling of NGC 7027 (\citealp{santander12}). The resulting best-fit of the Herschel/HIFI data is shown in Fig.~\ref{FNGC6302_2}, while the interferometric \doce\ and \trece\ \jdu\ synthetic maps are shown in Figs.~\ref{FNGC6302_3} and \ref{FNGC6302_4}. Because of the uncertainty in the HIFI data fluxes, which are of the order of 20-25\% (see \citealp{santander12}), we applied different free intensity scale factors to each of the model spectrum transitions (i.e. the factor to be applied to the intensity of the modelled profile to account for computational errors, and to be considered as the conservative error of the intensity of our model). In the best-fit model, the deviation of these factors from unity is within 30\% (i.e. the factors range from 0.77 to 1.3). Similar uncertainties are expected in the interferometric observations by \cite{dinhvtrung08}, with the additional problem that about 20\% of the flux is lost in those observations in a hardly predictable manner. Our model fits these maps fairly well, with differences between model calculations and data within observational uncertainties, although some strong observational variations from channel to channel could not be reproduced by our model.

The systemic velocity of the nebula is V$_\mathrm{LSR}$=-31~\kms. The relative abundance of the best-fit is 1.5$\times$10$^{-4}$ for \doce\ and 2.5$\times$10$^{-5}$ for \trece. The derived \doce/\trece\ abundance ratio may appear to be low, but we note that it is not unusual among post-AGB nebulae, particularly around O-rich stars \citep[see the thorough discussion in][]{bujarrabal13b}.

The results are summarised in Tables 1, 2, and 3, along with a conservative estimate of the uncertainties. These were computed by varying a given parameter while leaving the others unchanged, so that the deviation from the data remained within the aforementioned 30\%. For the velocities, the uncertainty was derived by eye, by varying the velocity  until the model did no longer provide a good fit. Note that, given the lack of information on the geometry of the nebula (because of the poor spatial resolution of the observations) and the main motivation of the model, we did not consider uncertainties on the geometry parameters.

\subsection{Discussion}

Channel-to-channel variations in the interferometric data of NGC 6302 show that the real nebula must be full of clumps, matter condensations and voids. Our model, however, provides an approximate description of the general properties and dynamics of the molecular envelope of this nebula and its characteristic physical conditions. 

The densities and temperatures of the different structures of our model are in general compatible with previous results by \citeauthor{dinhvtrung08} and \citeauthor{bujarrabal11} In particular, the temperature of the fastest blue-shifted structures (referred to as fast winds by \citeauthor{bujarrabal11}) is slightly higher (50-80 K compared to 40 K), and the densities somewhat lower ($\sim$10$^4$ cm$^{-3}$ compared to $\sim$10$^5$ cm$^{-3}$ ) than found by these authors. The density of the inner regions of the torus is of the order of a few 10$^5$ cm$^{-3}$, however, in good agreement with the findings by \cite{bujarrabal11}. 

No signs of shock-accelerated structures are apparent in our modelling, other than the fingers, which lie outside the plane of the torus and move away from the star, reaching velocities of up to 28 and 38 \kms. Note that although these structures might be at a different position along the same line of sight, the velocities are derived from the interferometric maps and thus are well constrained. The fingers seem to show slightly higher velocities (1-2 \kms) in their inner, hotter regions than in the neighbouring boundaries with the cooler, outer part of the same fingers. This might be indicative of a slow shock developing through the fingers, although there is not enough evidence for this assertion. The fingers share the main properties of the rest of the nebula at optical wavelengths: their projected expansion velocity is similar to the observed expansion in the same region at optical wavelengths (\citealp{costa14}), while their densities are similar to those found in the ionized region by \cite{rauber14}. To some extent, the fingers resemble the cometary tails found in the Helix (\citealp{odell96}), with a filamentary structure and a shielded region undergoing erosion.

Modelling low-$J$ and high-$J$ transitions together has allowed us to better characterise the physical conditions of the molecular envelope of NGC 6302. We derive a mass of 0.08 M$_\odot$ for the torus and 0.03 M$_\odot$ for the blobs and fingers, giving a total mass for the CO-emitting region of 0.11 M$_\odot$. This result is completely at odds with the mass computed by \cite{peretto07}, but in full agreement with the $\sim$0.1 M$_\odot$ found by \cite{dinhvtrung08}. Note that these authors used a simplified method to estimate the total mass from the total emission of a low-$J$ \trece\ line, which is expected to be optically thin and only slightly dependent on the excitation state.

\subsection{Summary}

We have built a model of the molecular envelope of NGC 6302 to investigate its physical and excitation conditions. The large-scale 3-D structure of our model molecular envelope was determined mostly from the direct interpretation of the interferometric maps and high-$J$ line profiles. Provided a simple velocity pattern, as typically observed in PNe, this structure is the simplest one that can account for all the observations at hand.

The nebula was modelled as a broken ring with an inner, hotter region and an extended, colder, outer one, together with a series of blobs and fingers expanding outwards from the plane of the ring. The velocity follows a ballistic expansion pattern except in the thinner structures, which have single, constant values of the velocity. The densities are in the range of 10$^4$~cm$^{-3}$ and 10$^5$~cm$^{-3}$ for the fingers and ring, while the temperatures range from a few tens of K in the outer regions of the ring to 300 K in the inner, exposed region. The mass of the molecular envelope is 0.11 M$_\odot$.

Our model provides a fair fit to the data. It is clear, however, that the intrinsic, small-scale structure of the nebula must be full of clumps and voids, as sudden channel-to-channel variations, which our model fails to reproduce, seem to imply in the interferometric maps. 

Finally, we stress that complex 3-D structures and dynamics can he modelled with a reasonable effort because {\tt shapemol} and its implementation within {\tt SHAPE} is fast and flexible.

\newpage

\begin{acknowledgements}
This work was partially supported by grant ``UNAM-PAPIIT 100410 and 101014'' and by the Spanish MICINN within the program CONSOLIDER INGENIO 2010, under grant ``Molecular Astrophysics: The Herschel and ALMA Era, ASTROMOL'' (ref.: CSD2009-00038).
\end{acknowledgements}

%\begin{thebibliography}{}

\bibliographystyle{aa}
%\bibliography{/home/santander/Research/papers/shapemol/msantander}
\bibliography{msantander}

%\end{thebibliography}

\newpage
\appendix

\section{Effects of vibrational transitions on the emission of
  considered rotational transitions} 

In our general calculations, we only considered rotational levels
within the ground-vibrational state, assuming that vibrational
transitions have little efect on the emission of the relevant $v$=0
rotational lines. To validate this assumption, we performed
calculations in some relevant cases, also including the $v$=1
rotational levels (40 additional levels) and an IR
continuum source. The IR source was always assumed to be central and
relatively small compared to the distance at which the studied point is
placed in the nebula (both assumptions are obviously not relevant to
our goals). The IR spectrum is assumed to be that of a black body for a
given temperature of about 500 - 10000 K, whose total size is calculated
to yield total IR intensities at 4.7 microns equal to the flux actually
observed in the studied nebulae (taking into account the distance to
them). This IR source can be a compact hot-dust region or a stellar
surface in cool stars.  We verified that the assumption of a black-body radiation spectrum and its temperature have negligible effects on
the final results, the relevant parameters being (as expected) the IR
source intensity and the distance between this source and the
considered element of molecule-rich gas. We also verified that in
all the studied cases the collisional vibrational excitation has
negligible effects on the excitation of the $v$=0 levels.

We considered three different model nebulae (see definitions of the
parameters in Sect.~\ref{shapemol}): 

A) A case similar to conditions expected in the Red Rectangle (a
post-AGB source in which most molecule-rich gas is placed in a rotating
disk, \citealp{bujarrabal13}): Distance to the object, $D$ = 710 pc,
F(4.7$\mu$) = 100 Jy, $V$ = 1.7 \kms, $r$ = 10$^{16}$ cm, relevant
densities 10$^{5}$ -- 10$^{6}$ cm${-3}$. Since in the Red Rectangle
rotation dominates dynamics, we have assumed a low value of $\epsilon$,
0.1 to simulate absorption by a long path between the centre and the
considered point. We recall that the LVG approach is not accurate in
the case of rotation, but the results of our tests give
an idea of the effects of the IR continuum in this interesting example. 

B) A case similar to conditions expected in the young PN NGC\,7027
(\citealp{santander12}): $D$ = 1000 pc, F(4.7$\mu$) = 50
Jy, $V$ = 15 \kms, $r$ = 10$^{17}$ cm, relevant densities $\sim$
10$^{4}$ -- 10$^{5}$cm${-3}$. Since in this source the velocity
gradients are often large, we took $\epsilon$ = 1, but we also present
calculations for a very low velocity gradient $\epsilon$ = 0.1 to see
possible effects due to this change.

C) A case similar to conditions expected in the PPNe CRL\,618
(\citealp{soria13}, \citealp{sanchez04}): $D$ = 1000
pc, F(4.7$\mu$) = 15 Jy, $V$ = 15 \kms, $r$ = 10$^{16}$ cm, relevant
densities $\sim$ 5 10$^{6}$ cm${-3}$. Some components of this source
(e.g.\ the very fast outflow) show a strong velocity gradient, but in
others (the fossil AGB envelope) the velocity increases slowly, so in
this case we also considered $\epsilon$ = 1, 0.1. 

We mostly calculated the intensity for a $J$=16--15 transition,
which has been widely observed with Herschel/HIFI, since low-$J$
transitions are more easily thermalised and their emissivities are more
easy to describe. We present in Figs.\ A.1 to A.5 predictions for the
A, B, C cases of the Raighleigh-Jeans equivalent brightness temperature
of \doce\ \trece\ \jdu\ transition:
$$
T_R = \frac{k}{h \nu} [I - I({\rm BG})],
$$
where $I$ is the intensity at the line centre emitted in the direction
perpendicular to the radial one and $I({\rm BG})$ is the intensity of
the cosmic background at the line frequency. 

\begin{figure*}[!]
\center
\resizebox{15cm}{!}{\includegraphics{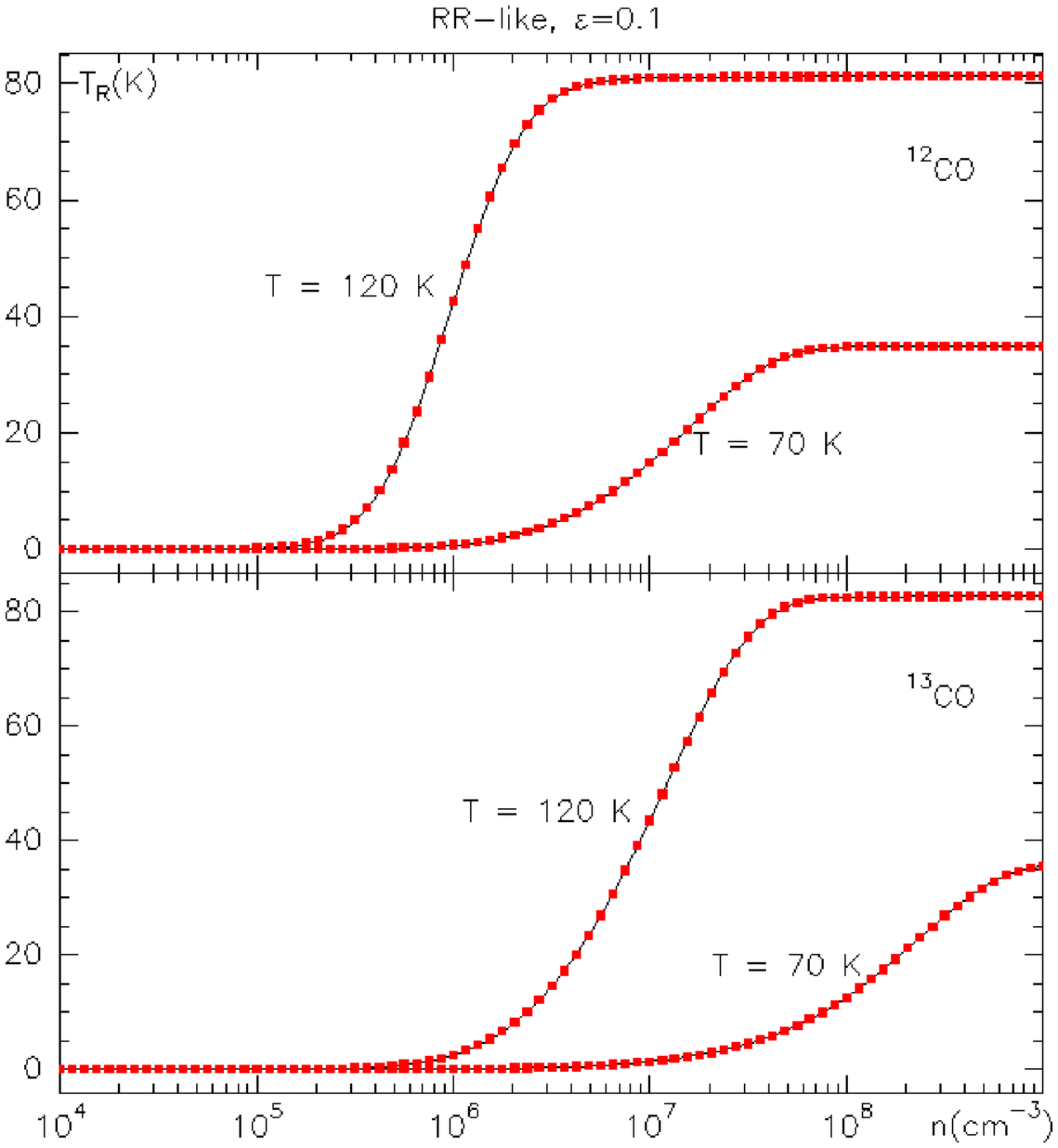}}
\caption{Model calculations of $T_\mathrm{R}$ as defined in Appendix A for the conditions corresponding to case A) in the text and an $\epsilon$ value of 0.1.}
\label{FA1}
\end{figure*}

\begin{figure*}[!]
\center
\resizebox{15cm}{!}{\includegraphics{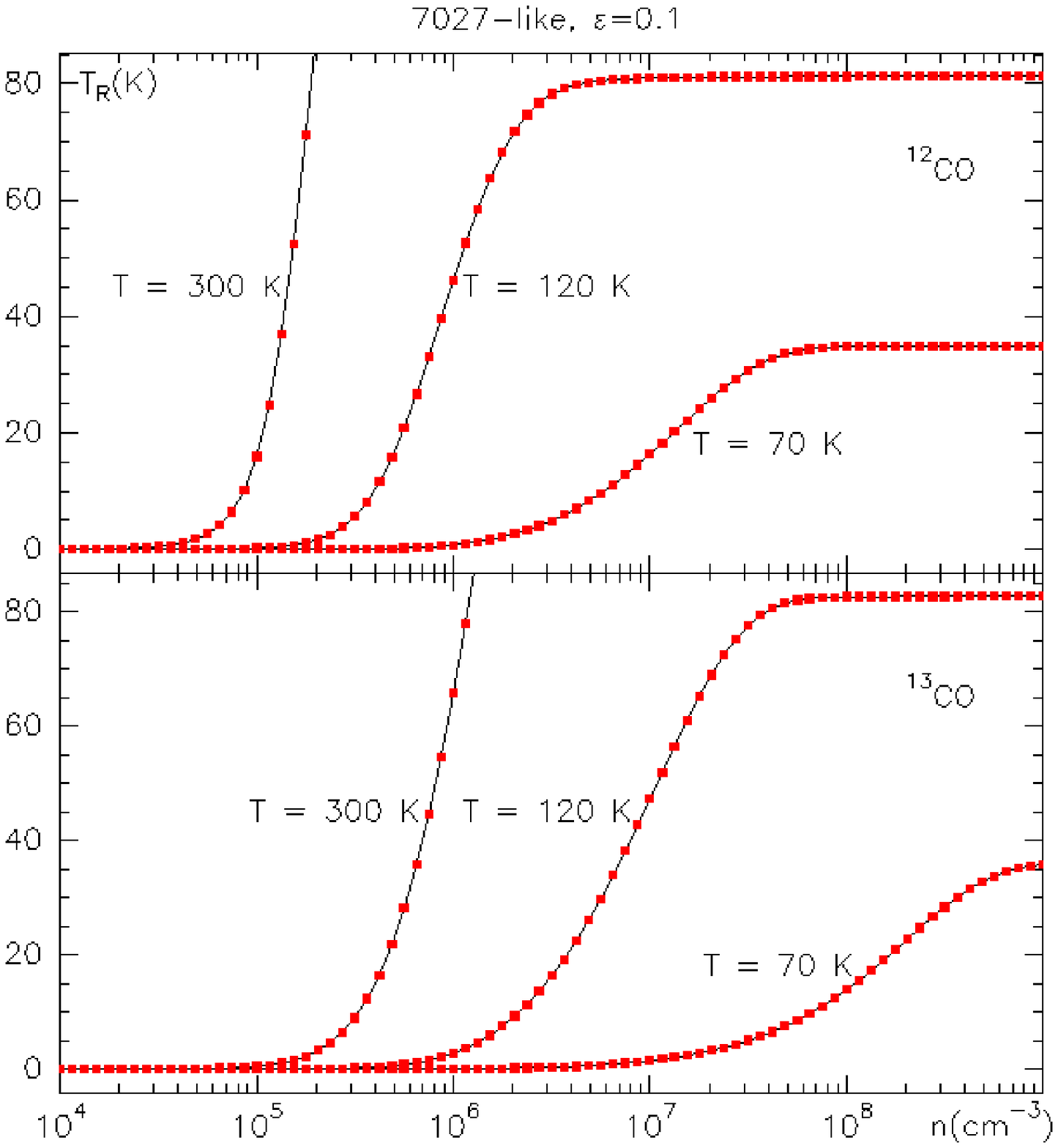}}
\caption{Model calculations of $T_\mathrm{R}$ as defined in Appendix A for the conditions corresponding to case B) in the text, for an $\epsilon$ value of 0.1.}
\label{FA2}
\end{figure*}

\begin{figure*}[!]
\center
\resizebox{15cm}{!}{\includegraphics{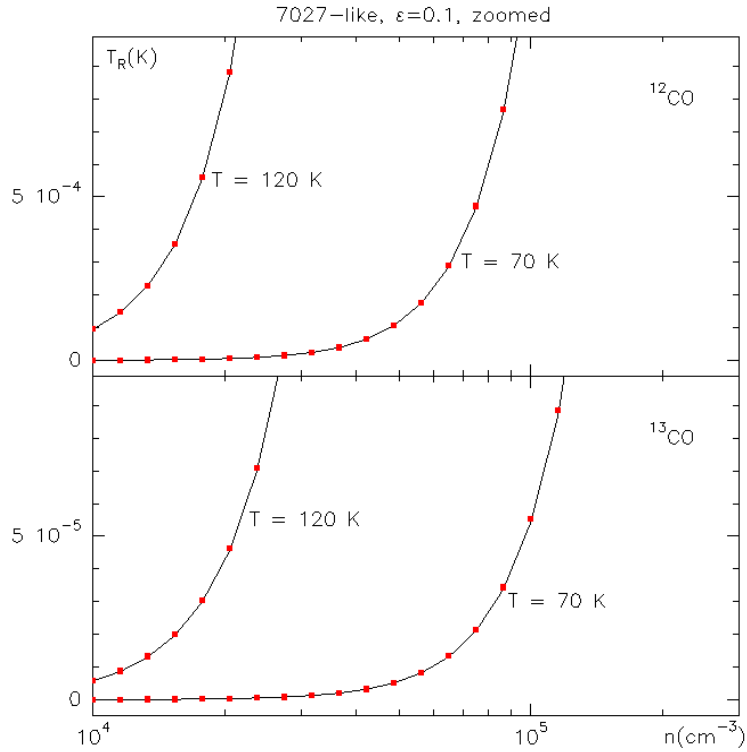}}
\caption{Close-up of case B) for an  $\epsilon$ value of 0.1.}
\label{FA3}
\end{figure*}

\begin{figure*}[!]
\center
\resizebox{15cm}{!}{\includegraphics{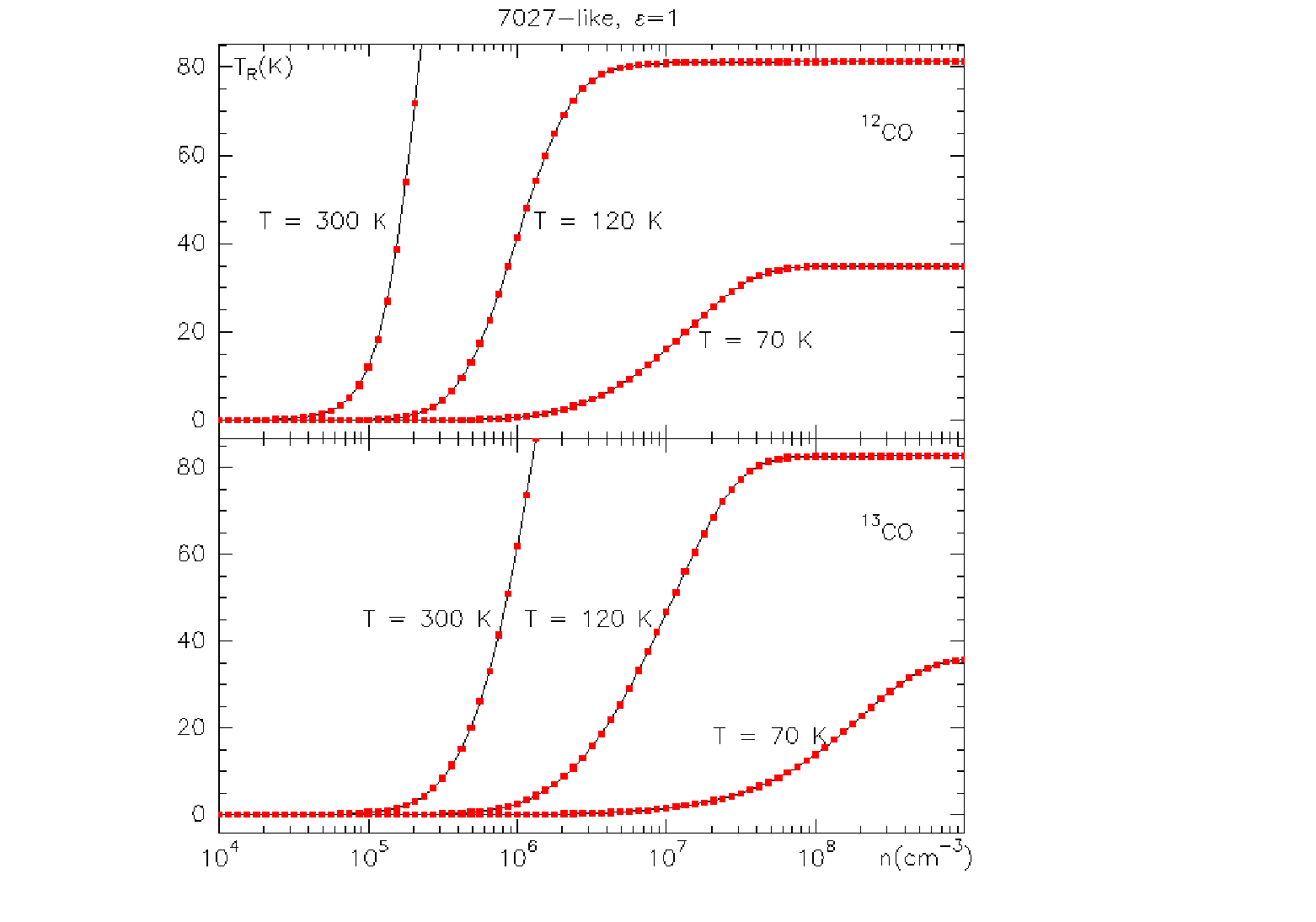}}
\caption{Model calculations of $T_\mathrm{R}$ as defined in Appendix A for the conditions corresponding to case B) in the text, for an $\epsilon$ value of 1.}
\label{FA4}
\end{figure*}

\begin{figure*}[!]
\center
\resizebox{15cm}{!}{\includegraphics{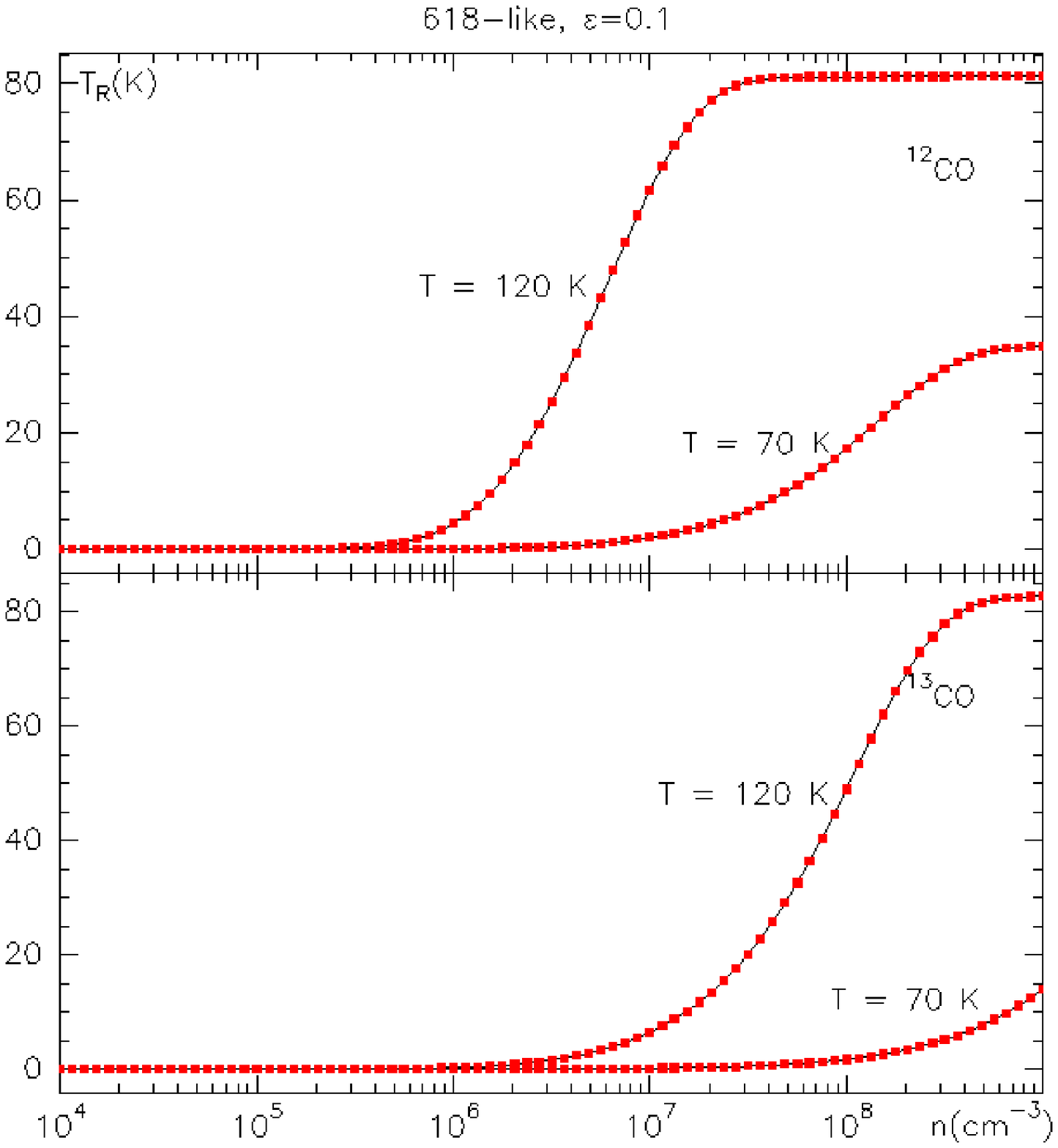}}
\caption{Model calculations of $T_\mathrm{R}$ as defined in Appendix A for the conditions corresponding to case C) in the text, for an $\epsilon$ value of 0.1.}
\label{FA5}
\end{figure*}

\begin{figure*}[!]
\center
\resizebox{15cm}{!}{\includegraphics{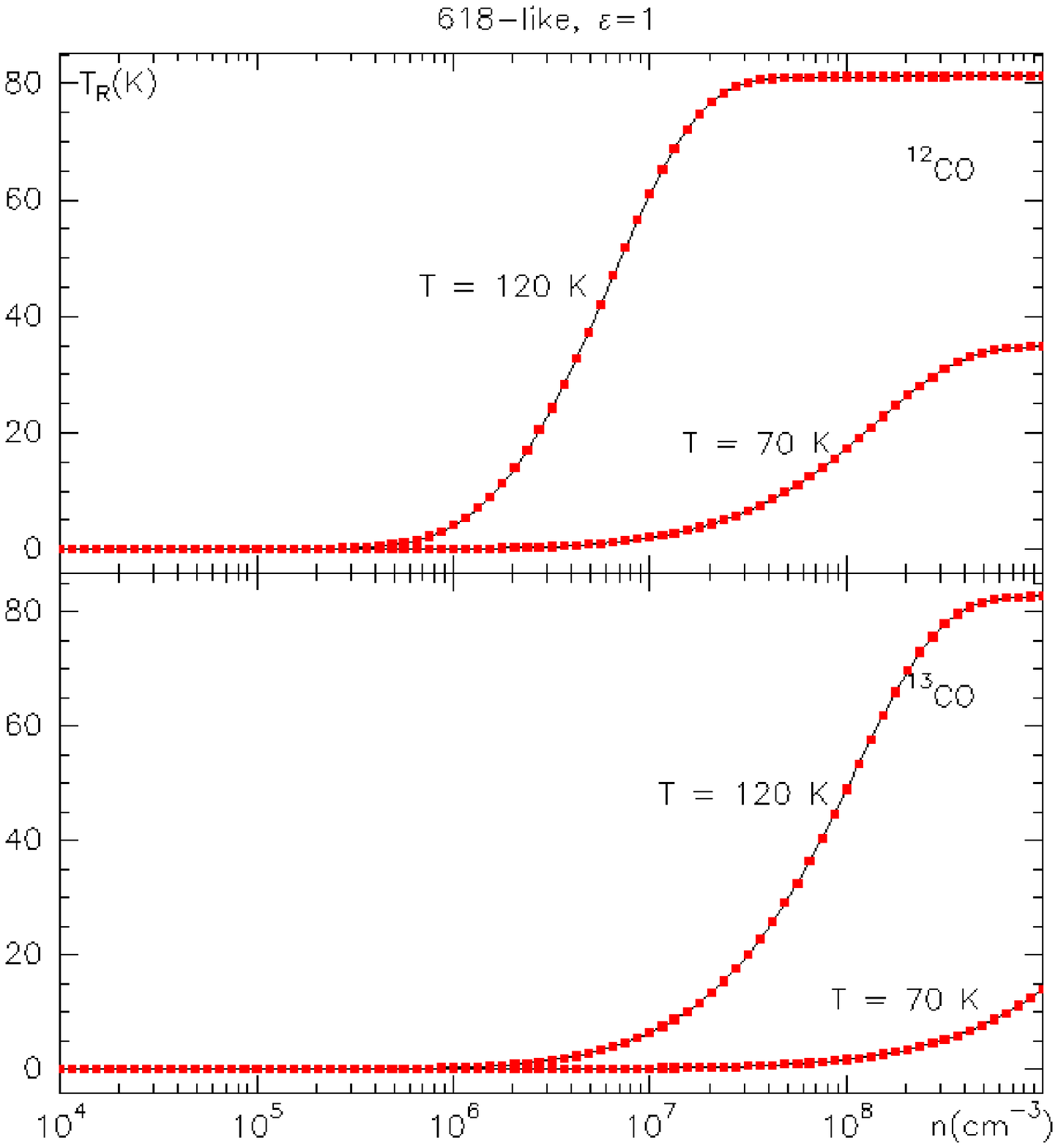}}
\caption{Model calculations of $T_\mathrm{R}$ as defined in Appendix A for the conditions corresponding to case C) in the text, for an $\epsilon$ value of 1.}
\label{FA6}
\end{figure*}

\begin{figure*}[!]
\center
\resizebox{15cm}{!}{\includegraphics{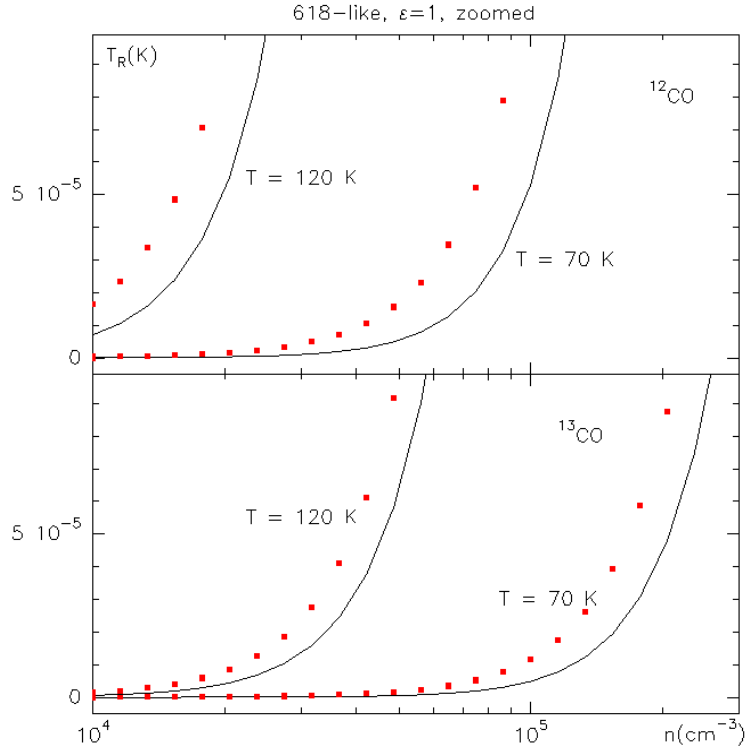}}
\caption{Close-up of case C) for an  $\epsilon$ value of 1. Deviations from the model are only significant for densities much lower than those found in the real nebula (see references in text).}
\label{FA7}
\end{figure*}

\end{document}